\documentclass[namedreferences]{solarphysics}
\usepackage[hyperref,optionalrh]{spr-sola-addons}
\usepackage{graphicx}                    
\usepackage{multirow}
\usepackage{color}                       
\usepackage{breakurl}                    

\usepackage{rotating}
\usepackage{caption}

\def\arcsec{\hbox{$^{\prime\prime}$}}

\def\my{Lo14}

 \newcommand{\B}[1]{{\color{black}{#1}}}

\pdfoutput=1

\begin{document}

\begin{article}

\begin{opening}

\title{Triggering an eruptive flare by emerging flux in a solar active-region complex}

\author{Rohan E. Louis$^{1}$\sep
        Bernhard Kliem$^{2,3}$\sep
        B. Ravindra$^{4}$\sep
        Georgios Chintzoglou$^{3}$
       }
\runningauthor{Louis {\it et al.}}
\runningtitle{Eruptive solar flare triggered by emerging flux}

\institute{$^{1}$ Leibniz-Institut f\"ur Astrophysik Potsdam (AIP),
	  An der Sternwarte 16, 14482 Potsdam, Germany
          email: \url{rlouis@aip.de}\\
	  $^{2}$ Institut f\"ur Physik und Astronomie,
          Universit\"at Potsdam, Karl-Liebknecht-Str. 24-25,
          14476 Potsdam, Germany\\
          $^{3}$ School of Physics, Astronomy and Computational Sciences,
          George Mason University, Fairfax, VA 22030, USA \\
          $^{4}$ Indian Institute of Astrophysics,
          Koramangala, Bengaluru 560034, India 
          }

\date{Received 12 February 2015 / Accepted ...}

\begin{abstract}
A flare and fast coronal mass ejection \B{originated} between 
solar active regions NOAA 11514 and 11515 on July~1, 2012 
in response to flux emergence in front of the leading sunspot 
of the trailing region 11515. Analyzing the evolution of the 
photospheric magnetic flux and the coronal structure, we find 
that the flux emergence triggered the eruption by interaction 
with overlying flux in a non-standard way. The new flux 
neither had the opposite orientation nor a location near the 
polarity inversion line, which are favorable for strong 
reconnection with the arcade flux under which it emerged. 
Moreover, its flux content remained significantly smaller 
than that of the arcade ($\approx40$\,\%). However, a loop 
system rooted in the trailing active region ran in part under 
the arcade between the active regions, passing over the site 
of flux emergence. The reconnection with the emerging flux, 
leading to a series of jet emissions into the loop system, 
caused a strong but confined rise of the loop system. This 
lifted the arcade between the two active regions, weakening 
its downward tension force and thus destabilizing the 
considerably sheared flux under the arcade. The complex event was 
also associated with supporting precursor activity in an enhanced
network near the active regions, acting on the large-scale overlying
flux, and with two simultaneous confined flares within the active
regions.
\end{abstract}


\end{opening}

\section{Introduction}
\label{intro}

Eruptive solar flares arise from a disruption of the coronal magnetic field,
which evolves into a coronal mass ejection (CME). In many cases a filament or
prominence eruption is associated. The flare source region comprises highly
sheared magnetic fields in the corona that overlie a photospheric polarity
inversion line (PIL)
\citep{1968SoPh....5..187M,1990ApJS...73..159H,2006ApJ...649..490W}. These
locations inevitably show a filament channel in the chromosphere and often
contain a filament or prominence in the corona above 
\citep{1998SoPh..182..107M}. It is widely (but not universally) accepted 
that a filament represents a weakly twisted magnetic flux rope holding the 
cool material \citep{2010SSRv..151..333M}. The formation and instability 
of a flux rope is a key element of storage-and-release eruption 
models \citep{1978SoPh...59..115V, 1995ApJ...446..377F, 2003ApJ...595.1231A,
2006ApJ...637L..65G, 2006ApJ...641..577M, 2006PhRvL..96y5002K}.

Eruptions can be triggered through the emergence of flux as well \B{as} through
flux cancellation. In the case of the former, the eruption occurs when
newly emerged magnetic fields appear in a region of pre-existing flux\B{, 
often in or in close proximity to a filament channel}
\citep[e.g.,][]{1972SoPh...25..141R,1982AdSpR...2...39M,1993A&A...271..292D,
1995JGR...100.3355F,2000PASJ...52..465L,2004ApJ...616..578S,
2009AdSpR..43..739S, 2012ApJ...748...77S}. This can lead to the formation of
a magnetic flux rope through bodily emergence
\citep{1996SoPh..167..217L, 2005ApJ...622.1275L}\B{, by} reconnection within
an emerged magnetic arcade \citep{2004ApJ...610..588M, 2010A&A...514A..56A},
\B{or by reconnection with the pre-existing flux \citep{2012ApJ...760...31K}.}
A number of variants for the triggering of eruptions by emerging flux have 
been proposed (see Section~\ref{s:discussion} for details).
According to nearly all of them, the presence of pre-existing flux,
orientated in a direction \B{that} facilitate\B{s} reconnection with the emerging flux,
is crucial to the eruption of the coronal flux rope
\cite[e.g.,][]{2000ApJ...545..524C, 2008A&A...492L..35A, 2012ApJ...760...31K, 
2014ApJ...787...46L}. Otherwise, a specific, narrow range of locations 
\citep{2008ChA&A..32...56X} or a large amount of emerging flux 
\citep{2014ApJ...796...44K} is required. 
\cite{2001JGR...10625053L} argued on general grounds that while the 
emergence of new flux can cause a loss of equilibrium, there is no simple, 
universal relation between the orientation of the emerging flux and the 
likelihood of an eruption, due to the complexity of the equilibrium 
surface in the multidimensional parameter space of the configuration.

Flux cancellation is also an important mechanism for triggering solar flares
\citep{1989SoPh..121..197L, 2010A&A...521A..49S, 2011A&A...526A...2G,
2012ApJ...759..105S, 2013SoPh..283..429B}. When canceling photospheric 
magnetic flux patches lie at the base of sheared coronal flux, then the 
cancellation is associated with magnetic reconnection low in the atmosphere\B{. 
This gradually forms a filament channel in the photosphere and chromosphere and} 
a flux rope above the PIL where the cancellation occurs 
\citep{1989ApJ...343..971V, 1993SoPh..143..119W, 2006ApJ...641..577M, 
2010ApJ...708..314A, 2011ApJ...742L..27A}. The flux rope gradually rises as 
more flux is reconnected into it, implying a higher current flowing along 
the rope. When the rope becomes unstable and rises much faster, a vertical 
current sheet is formed underneath and becomes the site of fast ``flare'' 
reconnection \citep{2000JGR...105.2375L}.

In this article we employ multiwavelength and dual-viewpoint observations 
to analyze an eruptive flare that \B{originated} between two active regions (ARs) 
that formed a quadrupolar active-region complex. Both the emergence 
of new flux and flux cancellation occurred under the erupting middle flux 
lobe. The flux cancellation was weak, leaving the flux emergence as the 
primary candidate for the triggering of the eruption. However, the flux 
emergence did not correspond to any of the standard scenarios for such 
triggering: the direction and emergence site of the new flux were not 
favorable for strong reconnection with the pre-existing erupting flux, and its 
magnitude was only moderate. Additionally, the eruption was preceded 
by a complex sequence of events in the vicinity of the AR complex and 
accompanied by confined flares within the two ARs. The data analysis 
yields a suggestion for a non-standard triggering of the eruption by 
the emerging flux. We also discuss this eruption from a quadrupolar 
source region in relation to the breakout model 
\citep{2008ApJ...683.1192L, 2012ApJ...760...81K}.
\B{A further eruptive flare in the region of flux emergence, following 
about 19~hr later \B{and likely due to flux cancellation}, was analyzed 
in \citeauthor{2014A&A...562A.110L} (\citeyear{2014A&A...562A.110L}, 
hereafter referred to as \my)}.

\begin{figure}
\centerline{
\includegraphics[angle=0,width = 0.7\textwidth,clip=]{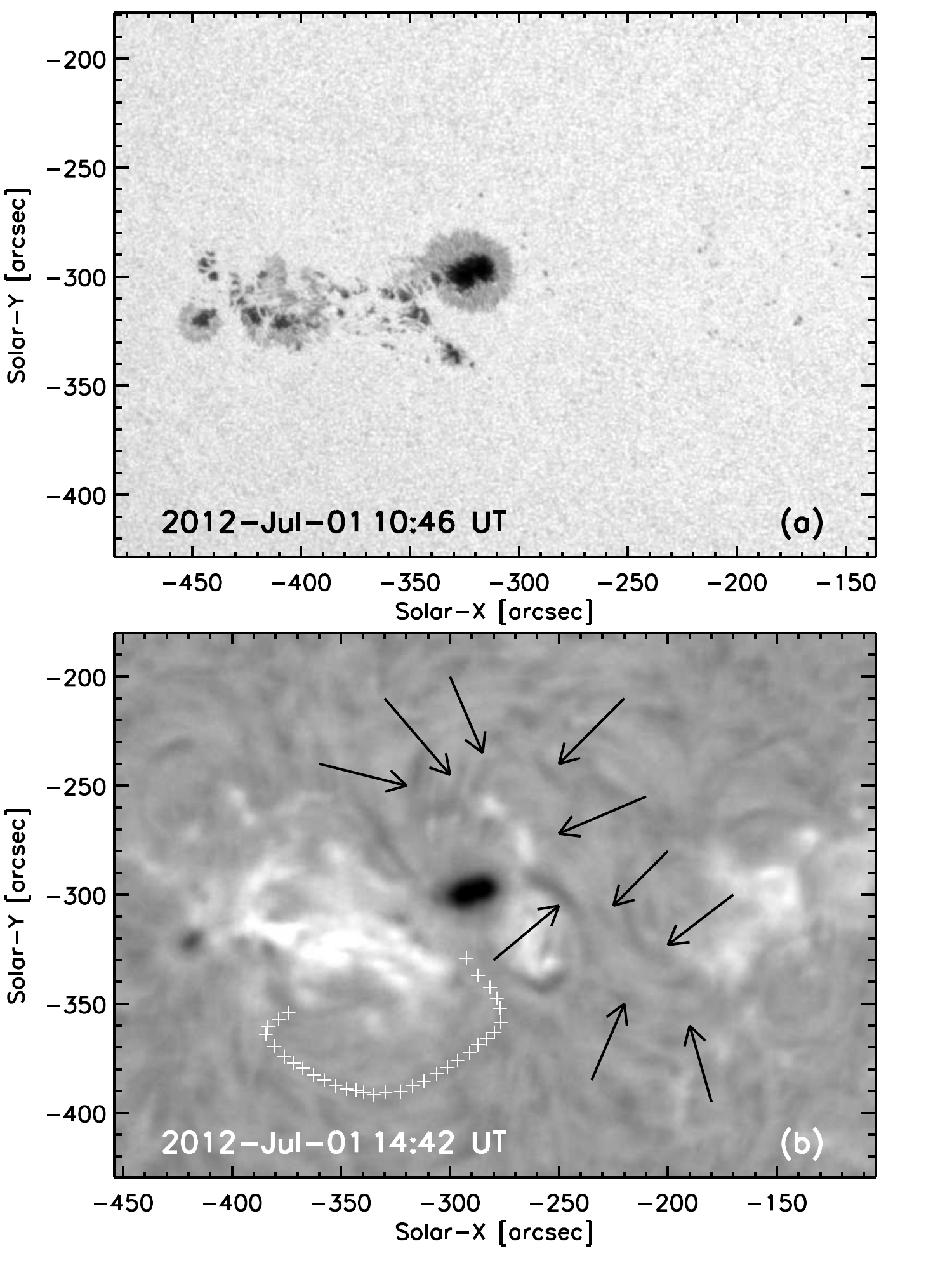}}
\caption{Overview images of the AR complex. 
(a) HMI continuum intensity; (b) GONG H$\alpha$ 
from Big Bear Solar Observatory (H$\alpha$ images from other 
observatories near the time of the other \B{panels in 
Figures~\ref{f:overview1} and \ref{f:overview2}} are less sharp). 
\B{Arrows in the bottom panel mark fibrils which indicate the forming 
filament channel involved in the eruption} {\color{black}while the plus symbols 
\B{indicate the overall alignment of H$\alpha$ fibrils} south of AR 11515 
with the loop system seen in the AIA 171 \B{and 193}~{\AA} images 
(Figure~\ref{f:overview2}(c--d)),} \B{similar to the absorbing 
material seen in He~{\sc ii}~304~{\AA} (Figure~\ref{f:overview2}(b))}.
An animation of HMI continuum and magnetogram and AIA 171, 304, and 
1700~{\AA} images during July~1, 2012 is provided as online material.}
\label{f:overview1}
\end{figure}

\begin{figure}
\centerline{
\includegraphics[angle=270,width = \textwidth,clip=]{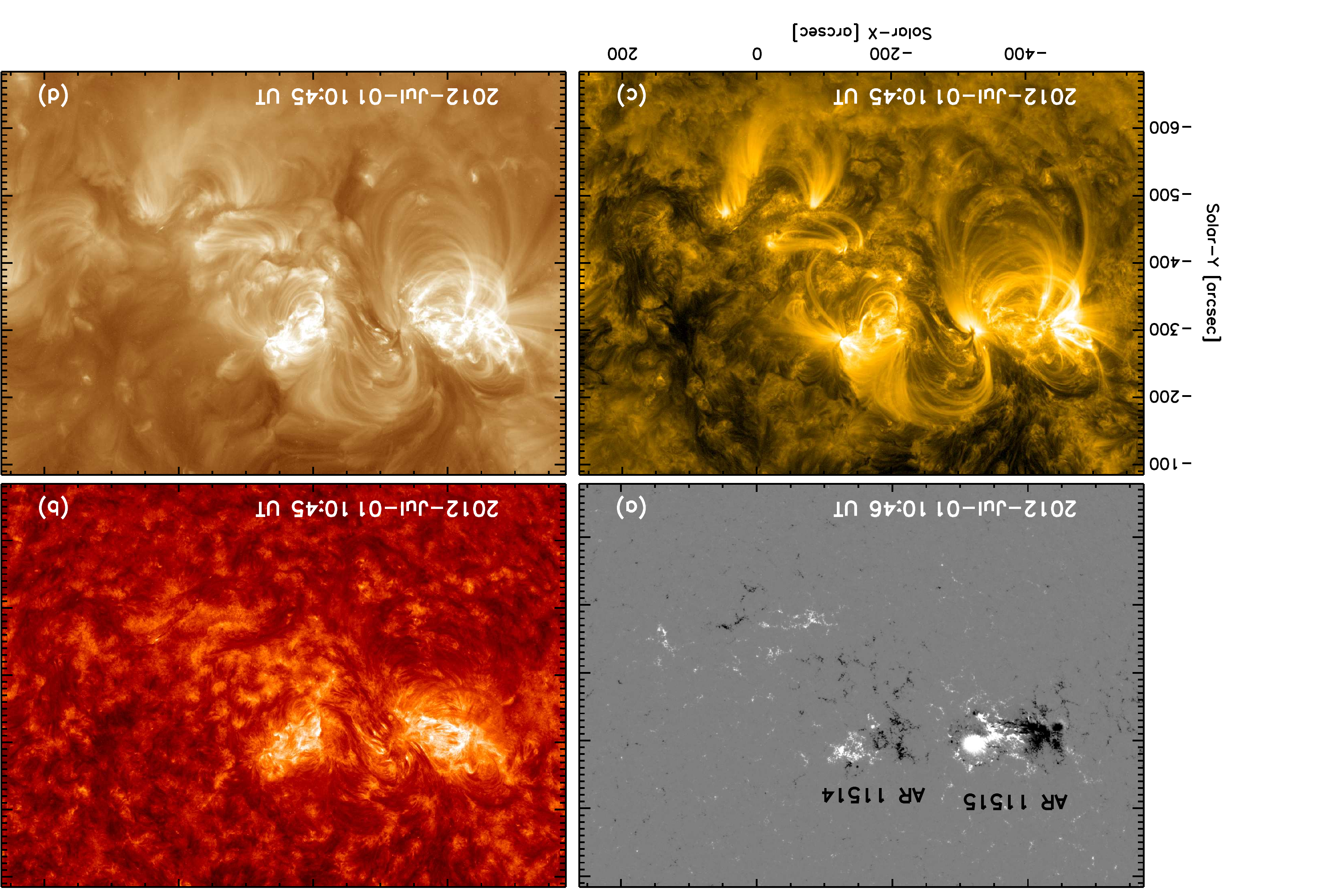}}
\caption{Continuation of Figure~\ref{f:overview1} showing \B{(a)} the HMI LOS magnetogram 
and the chromospheric and coronal structure in AIA images at (b) 304~{\AA}, 
(c) 171~{\AA}, and (d) 193~{\AA}. \B{The enhanced network area southwest 
of the AR complex, which contains the S-shaped filament centered at 
solar--$(x,y)=(0,-480)$,
is included here.}}\vspace{-10pt}
\label{f:overview2}
\end{figure}

\section{Observations and data processing}
\label{data}

We used full-disk solar observations from the Solar
Dynamics Observatory \citep[SDO;][]{2012SoPh..275....3P} that include data
from the Helioseismic and Magnetic Imager \citep[HMI,][]{2012SoPh..275..229S}
and the Atmospheric Imaging Assembly \citep[AIA,][]{2012SoPh..275...17L} for
the two ARs of interest. The HMI data set consists
of 4k$\times$4k-pixel images of continuum intensity and
line-of-sight (LOS) magnetograms derived from the photospheric Fe~{\sc i}~617.3~nm
line at a spatial sampling of 0.5\arcsec~pixel$^{-1}$.
The HMI data were processed in a similar manner as described in \my.
The AIA data consist of 4k$\times$4k-pixel images with a spatial
sampling of 0.6\arcsec~pixel$^{-1}$ in a range of UV and EUV bands.
In particular, we have analyzed the bands including He~{\sc ii}~304~{\AA},
Fe~{\sc ix}~171~\AA, Fe~{\sc xii}~193~\AA, Fe~{\sc xiv}~211~\AA,
Fe~{\sc xvi}~335~\AA,  Fe~{\sc xviii}~94~\AA, Fe~{\sc xxi}~131~\AA,
and the continuum at 1700~{\AA}.
The observations span a duration of 28~hr starting on 2012 July~1 with a
cadence of 12~sec for the AIA data and 45~sec for the HMI data.

We also used full-disk H$\alpha$ images from the Global Oscillations
Network Group \citep[GONG;][]{1996Sci...272.1284H,2011SPD....42.1745H}
recorded at Big Bear Solar Observatory.
The 2k$\times$2k images have a spatial sampling of about 1\arcsec~
with a cadence of 5~minutes and cover the passage of the \B{eruptive} flare.

Limb views of the event are provided by the SECCHI instrument package
\citep{2008SSRv..136...67H} onboard the STEREO Behind spacecraft. We analyzed
195 and 304~{\AA} data from the EUVI-B imager and data from the COR1-B 
coronagraph.

\section{Source region}
\label{s:source}

\begin{figure}
\centerline{
\includegraphics[angle=0,width = 0.7\textwidth,clip=]
{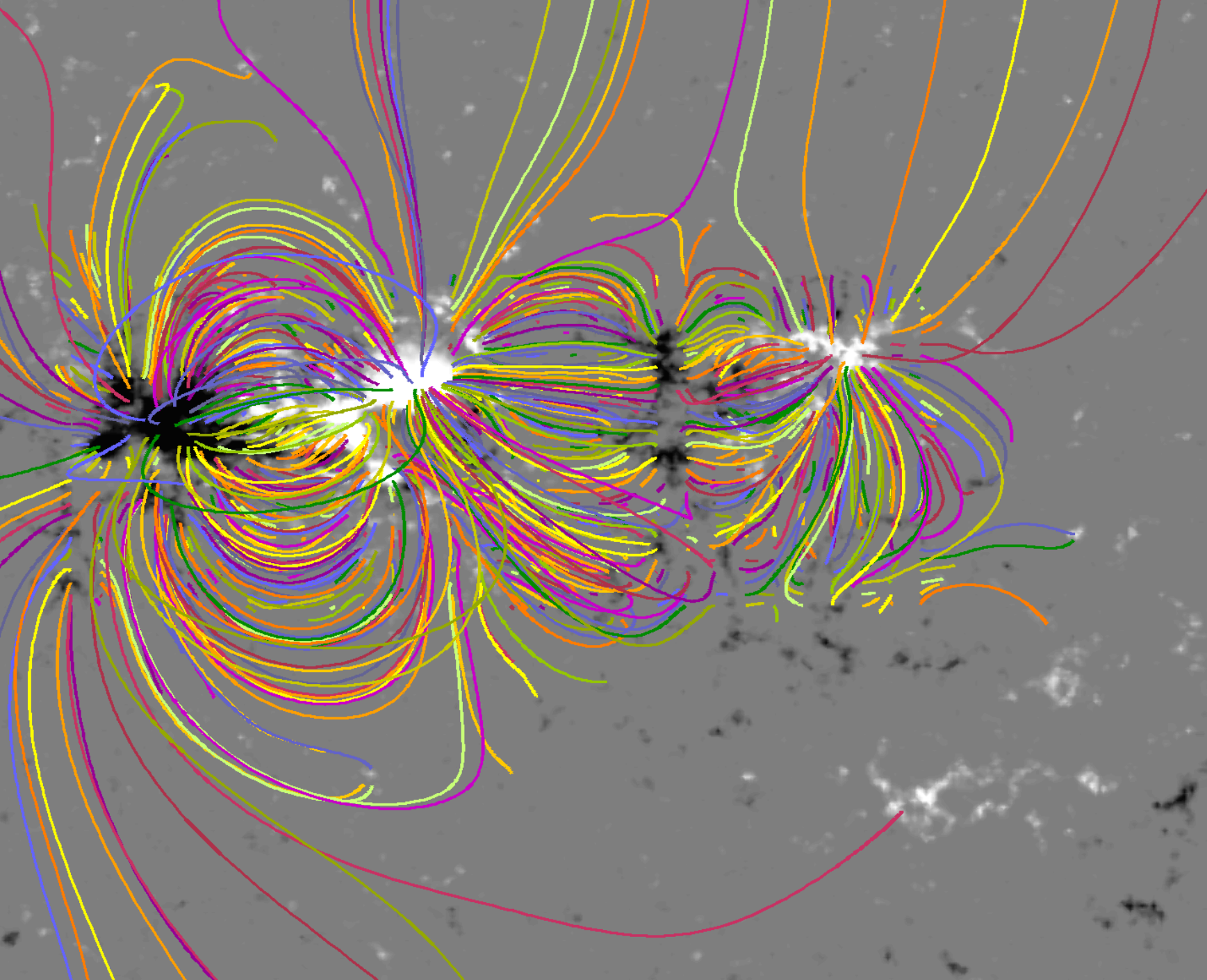}}
\caption{Field lines of a PFSS extrapolation using the HMI LOS magnetogram of 
ARs~11514 and 11515 and their environment at 15:24:45~UT patched into a synoptic 
full-Sun magnetogram. The field is computed using the method by 
\citet{1995JGR...100...19Z} with 330 spherical harmonics and the source 
surface located at $2.5~R_\odot$. The grayscale of the magnetogram display is 
saturated at $\mid$B$\mid=$500~G.}
\label{f:PFSS}
\end{figure}

NOAA ARs 11514 and 11515 appear at the east limb on June~26, 2012 as young 
regions still in their emergence phase. By the beginning of July~1, their 
centers are located at S15E09 and S16E24, respectively. The leading AR~11514 
is then already dispersing and spotless, while the following AR~11515 still 
develops strongly, with considerable rearrangements of its multiple spots. 
Both regions are essentially bipolar and follow Hale's law, so that they 
jointly form a quadrupolar configuration. {\color{black}AR~11515 contains 
far more flux, amounting to $2.6\times 10^{22}$~Mx, while the total magnetic flux
in AR~11514 is around $6.1\times 10^{21}$~Mx.}
Southwest of AR~11514, four areas of enhanced network with alternating 
polarity are located. Filaments have developed at all three resulting PILs, 
with the first two forming a joint, forward S-shaped filament, which becomes 
involved in the eruption studied here. The AR complex and adjacent 
enhanced network area are displayed at various wavelengths in 
Figures~\ref{f:overview1} and \ref{f:overview2}, and the accompanying 
animation shows their evolution during July~1. Their appearance in the 
AIA 211 and 335~{\AA} bands is very similar to the appearance at 193~{\AA}. 
It is also worth mentioning that AIA 211 and 193~{\AA} images provide a weak 
suggestion for the existence of a magnetic connection between AR~11515 and 
AR~11513 located in the other hemisphere at N15E12. Both AR~11515 and 11513 
are rather active during their disk passage, with much of temporally 
associated, possibly sympathetic activity occurring in the days around 
July~1--2.

\begin{figure}
\centerline{
\includegraphics[angle=0,width = 0.75\textwidth]{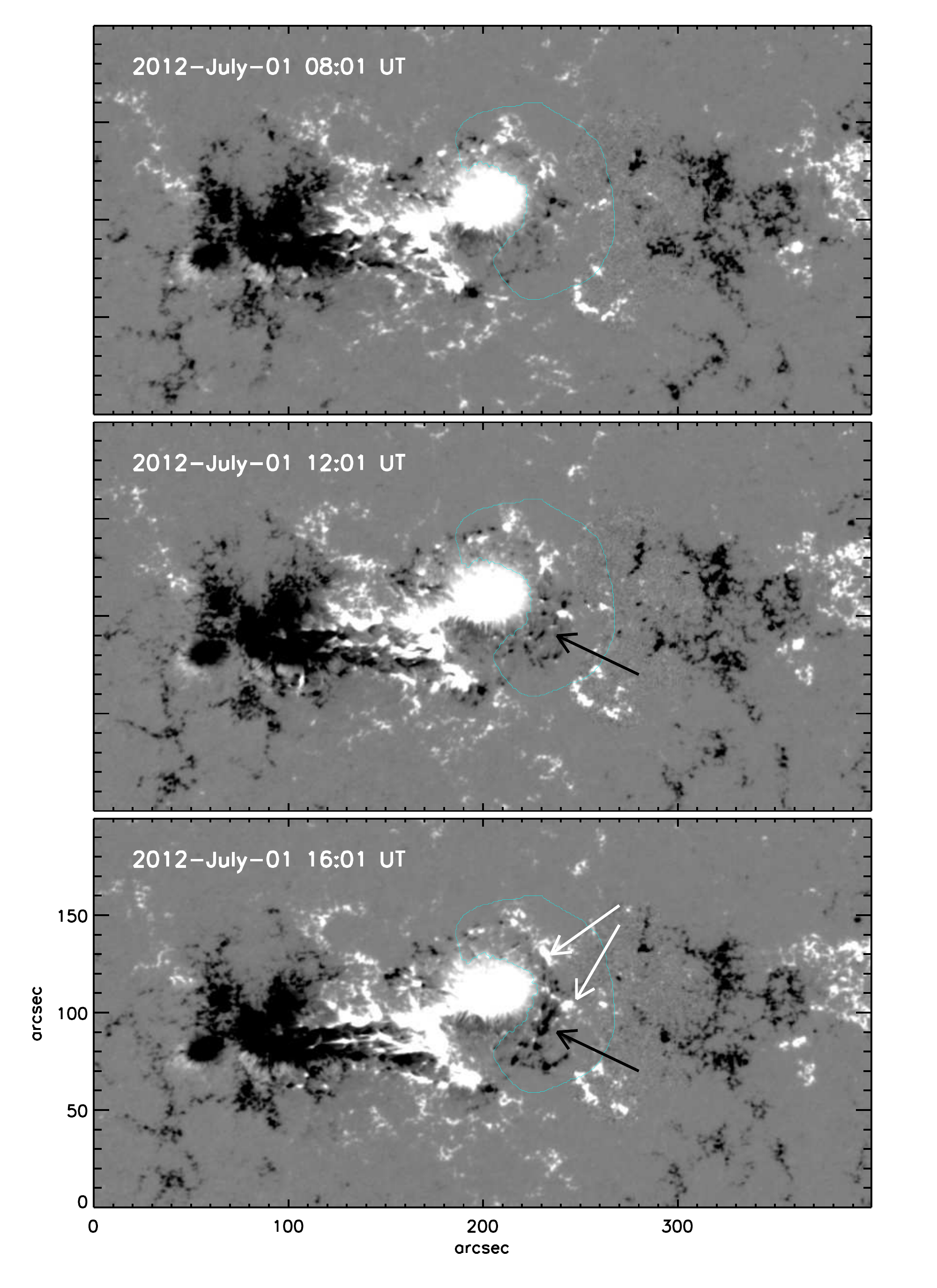}}
\caption{Emergence of new flux ahead of the LS in AR~11515. Mostly negative 
new flux can be seen, as indicated by the black arrows. Regions of increasing 
positive flux are marked by white arrows. The grayscale is saturated at 
$\mid$B$\mid=$500~G, except around the PIL between the ARs, where the 
flux has been enhanced by a factor of 4, to visualize the motion of
weak negative flux in the trailing part of AR~11514 toward the PIL in the 
final hours before the eruption. {\color{black}The cyan contour outlines the region of
emerging flux.}}
\label{f:emergence}
\end{figure}

\begin{figure}
\centerline{
\includegraphics[angle=180,width = 0.7\textwidth]{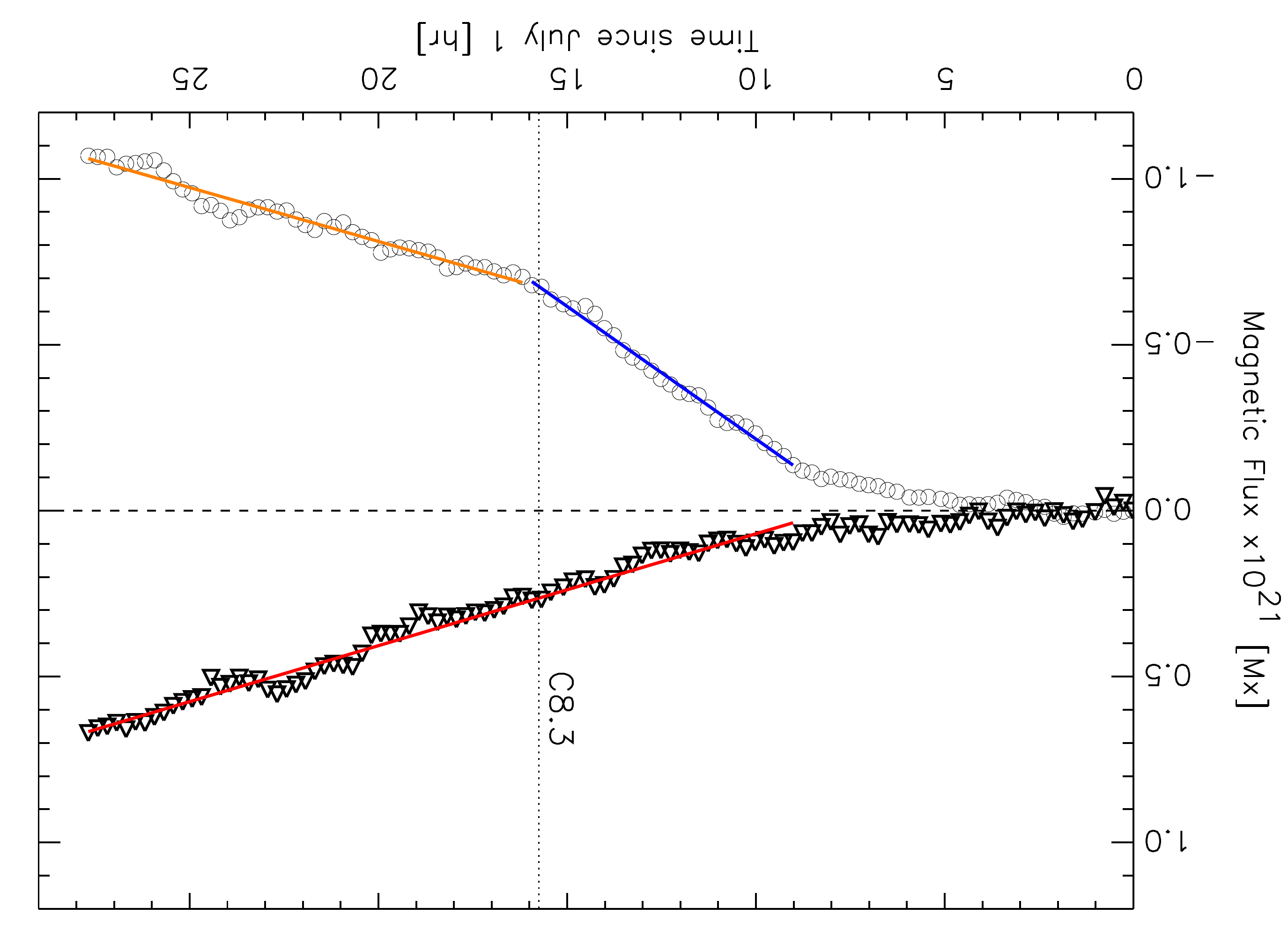}}
\vspace{-5pt}
\caption{Temporal evolution of magnetic flux in the emergence area ahead of
the LS {\color{black}(area enclosed by the cyan contour in Figure~\ref{f:emergence})} 
in AR~11515 over the course of 2012 July~1. The triangles 
and circles correspond to positive and negative flux, respectively. The 
flux shown in the plot is relative to the initial flux at 00:00~UT. The 
blue and orange lines represent linear fits to the observed values, before 
and after the occurrence of the C8.3 flare.}
\label{f:flux} 
\end{figure}

Figure~\ref{f:PFSS} shows field lines of a potential field source 
surface (PFSS) extrapolation of the AR complex immediately before 
the onset of the eruptive flare. One can clearly see that the two 
ARs are magnetically coupled: the excess positive flux in AR~11515, 
largely concentrated in the leading sunspot (LS), mostly connects to 
the excess negative flux in AR~11514, located in its trailing part. 
This yields the three-lobe magnetic configuration characteristic 
of a quadrupolar source region and agrees with the 
conclusion of {\my} that the (later splitting) LS has strong magnetic 
connections with AR~11514. The PFSS extrapolation also indicates 
magnetic connections of both ARs with the enhanced network area 
southwest of AR~11514, and north-ward pointing field lines starting 
in the LS, which most likely connect to AR~11513, a region dominated by 
negative flux. High-reaching field lines are also rooted in both 
the leading positive polarity of AR~11514 and the trailing 
negative polarity of AR~11515. These may indicate the existence 
of an overarching fourth lobe of flux, as required by the breakout 
model, or open flux. While the PFSS extrapolation is not conclusive 
in this regard, the EUV observations \B{described} below 
\B{(see Sections~\ref{ss:pre-flare}--\ref{ss:flare} and}
{\color{black}Figures~\ref{f:activation_AIA} and \ref{f:flare_AIA})} strongly suggest 
the existence of overarching flux, albeit with a different 
orientation. Moreover, the PFSS extrapolation indicates 
that the amount of overarching flux is considerably smaller 
than the flux in the three central lobes.

Beginning at about 05~UT on July~1 and continuing into the 
following day, new flux emerges in AR~11515 just in front 
\B{(westward)} of the LS, i.e. under the 
middle magnetic lobe of the AR complex, near the left edge 
of the lobe (Figures~\ref{f:emergence} and \ref{f:flux}). 
This is dominantly negative flux mixed with some small 
positive flux patches, thus forming a new PIL in front 
of the LS. It is difficult to discern with certainty 
where the corresponding positive flux emerges. Part of 
it must emerge within the preexisting scattered positive 
flux just outside the northwest periphery of the LS 
\B{(marked by a white arrow)}. The 
amount of this flux clearly increases during the emergence 
of the negative flux ahead of the LS (Figure~\ref{f:flux}). 
Additionally, a moderate positive flux patch grows on the 
western side just in front of the emerging negative flux 
\B{(marked by a second white arrow)}. 
It is possible, however, that part of the positive flux 
emerges in the moat of the LS, where it is hard to detect. 
Any coronal field lines connecting the LS and the emerging 
negative flux must have a direction rather similar to the 
overlying flux of the middle lobe of the AR complex. 
However, the field lines connecting the two polarities 
within the emerging flux in front of the LS have a 
considerable north-south component, forming \B{a small to moderate} angle with 
the overlying flux. \B{This} allows for the occurrence of 
magnetic reconnection\B{, although only with a moderate likelihood}. 
The minor part of the emerging 
flux rooted in the moderate positive flux patch on the 
western edge of the emergence area is directed oppositely 
to the overlying flux; this flux has a high likelihood 
for reconnection to occur. The relative amount 
of emerged flux by the time of the eruptive flare is estimated 
from the measurement of negative flux in Figure~\ref{f:flux} 
and from the amount of negative flux in the footpoint area 
of the middle flux lobe in the PFSS extrapolation 
to be $\sim40\,\%$.

Another major development in AR~11515 is the splitting of 
its LS, which becomes apparent after the eruptive flare 
studied here and is analyzed in \my.

In the EUV (see Figure~\ref{f:overview2} and the corresponding animation), 
both active regions are bright 
throughout the observation, with the usual multitude of small-scale 
brightenings and changing coronal loops reflecting the ongoing evolution 
of the photospheric flux distribution. Additionally, the region of new 
flux emergence is bright and highly variable. Before the eruptive flare, 
there are fewer coronal loops between the ARs, and they are mostly 
fainter than the loops within the ARs. Many of the inter-region loops 
are also distinguished by their clear reverse-S shape, indicative of 
magnetic shear, and by their higher temperature, as they are more 
pronounced in the ``warmer'' AIA bands (193, 211, and 335~\AA) than in 
the 171~{\AA} band. The higher temperature is seen already before 
the new flux emergence and is therefore indicative of the dissipation of 
enhanced currents in the sheared field. 

\begin{figure}
\centerline{
\includegraphics[width = \textwidth,clip=]{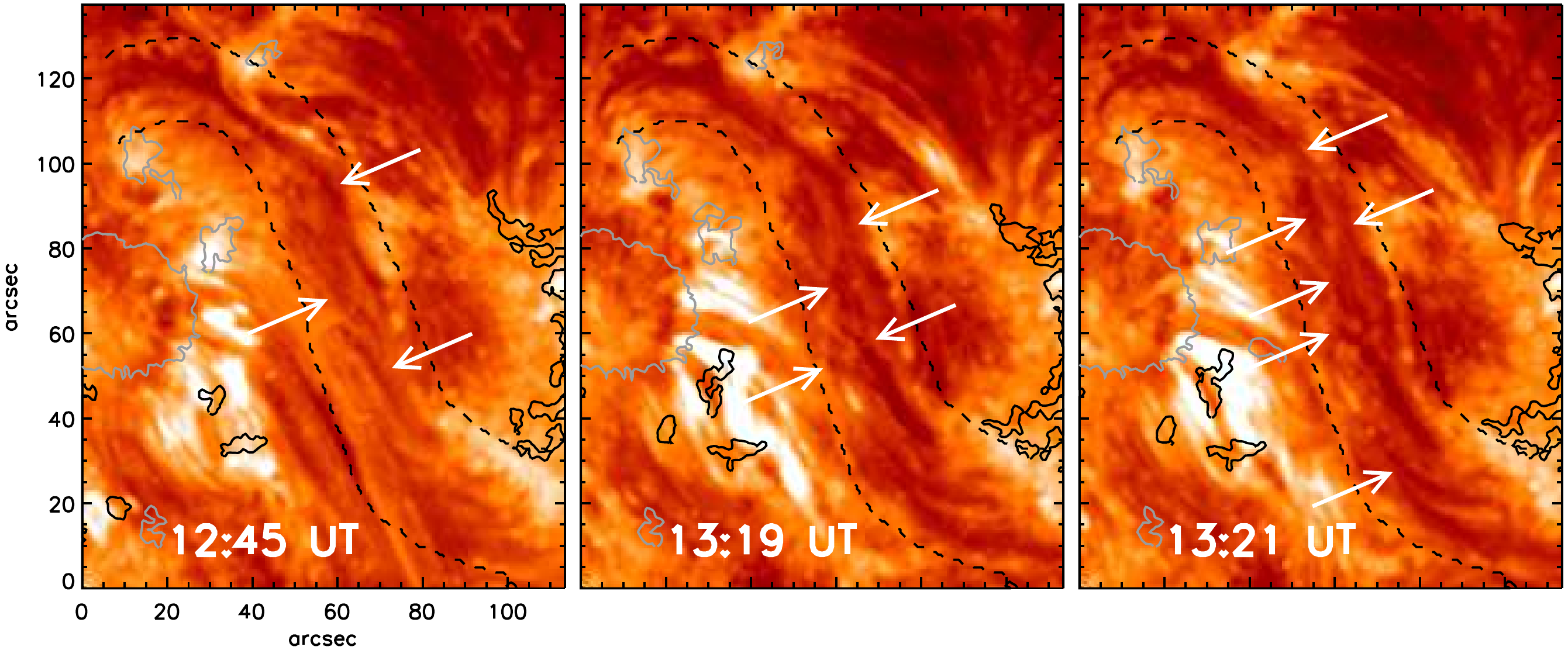}}
\caption{AIA 304~{\AA} images of the forming filament channel between the two 
ARs showing instances of apparently crossing threads in the middle of the 
channel (indicated by the arrows). {\color{black}\B{D}ashed lines
outline the filament channel. Gray (black) 
contours represent the LOS magnetic field of positive (negative) polarity.}}
\label{f:channel}
\end{figure}

The existence of considerable 
shear in the field between the ARs is also obvious from the strong 
alignment of the He~{\sc ii} threads at the PIL of the central field 
lobe, \B{shown in Figure~\ref{f:channel}}. A filament channel,
\B{indicated in the figure,} is obviously forming here; it extends with a 
reverse-S shape from the northern edge of the LS in AR~11515 to the southern 
edge of the trailing flux in AR~11514. \B{Its shear is left-handed, 
consistent with} the reverse-S shape of the overlying coronal loops. 
H$\alpha$-absorbing material can be seen to be arranged along the forming 
filament channel in elongated patches (Figure~\ref{f:overview1}). These 
patches are often changing \B{in} shape and position, with the trend to delineate 
the filament channel more completely by the end of the day.

Flows converging at the PIL are known to be essential for the formation 
of a filament channel \citep{1998SoPh..182..107M}. Since the photospheric 
field is weak in a broad strip around the PIL between the two ARs, an 
attempt to track magnetic features using the DAVE \B{algorithm} \citep{2006ApJ...646.1358S} 
did not yield a reliable map of the photospheric flow field around 
the PIL. However, the sequence of HMI magnetograms unambiguously shows that 
the narrow strip of weak, mostly negative flux at the \B{trailing} edge of AR~11514 
moves toward the PIL as a consequence of a strong expansion of the adjacent 
network cell from about 10~UT onwards (see Figure~\ref{f:emergence} and the 
AIA 1700~{\AA} data included in the online movie).

The detailed arrangement of the threads in the forming filament channel is 
highly variable. Overall, they indicate a highly sheared, left-handed arcade 
at all times, but at a number of instances several threads display a crossing 
pattern that might indicate the formation of a flux rope (see examples in 
Figure~\ref{f:channel}). Since this pattern does not persist, the actual 
formation of a flux rope at this location remains unclear.

Absorbing (filament) material is also observed in the AIA 304~{\AA} band 
\B{(Figure~\ref{f:overview2}(b))}
in the heterogeneous structures extending up to $\sim20^\circ$ south of 
AR~11515, which on a large scale are similar to the loops seen in the 
``warm'' bands at 171~{\AA}, 193~{\AA}, and higher 
\B{(Figure~\ref{f:overview2}(c,d))}. This cool material 
is also seen above the limb in EUVI-B 304~{\AA} images, with variable 
shape \B{(Figure~\ref{f:activation_EUVI}(a))}. The H$\alpha$ 
images show an arrangement of the fibrils in this 
area in a large-scale loop pattern that resembles the structure in the 
AIA 304~{\AA} band \B{(some of them are indicated by crosses in 
Figure~\ref{f:overview1}(b))}.

Finally, we note that right-handed helicity is indicated for the 
filaments in the enhanced network southwest of AR~11514 by the 
orientation of the absorbing He {\sc ii} threads, the shape of the 
brightenings during their later activation \B{(Section~\ref{ss:pre-flare})}, 
and by the forward-S shape of the western filament.

\section{Eruptive flare and associated pre- and post-flare activity}
\label{s:eruption}

Both ARs~11515 and 11513 display frequent subflaring and flaring activity 
throughout the days around 2012 July~1, some of which might be sympathetic 
activity. Therefore, it is difficult to identify with certainty the initial 
element in a chain of events leading to the eruptive C-class (C8.3) flare 
in the AR complex formed by ARs~11514 and 11515, which peaks at 15:45~UT. 
These activities occur in different places in and around the AR complex, 
indicating a highly complex structure of the magnetic field, even if a 
possible connection to AR~11513 on the other hemisphere is not taken 
into account. Only part of this complexity can be inferred from the 
EUV images in the present article.

The basic sequence of events consists of (1) two \B{relatively similar} 
activations of the 
loop system containing filament material south of AR~11515, which 
remain confined, (2) the eruptive C8.3 flare and fast CME, which 
originate at the PIL between the two ARs but involve a complex 
pattern of bright ribbons and dimming areas scattered over the 
AR complex and the neighboring area of enhanced network flux 
containing two filaments, and (3) related confined flaring 
activities \B{within} the \B{two ARs} in the peak and decay phases 
of the eruptive flare.

\subsection{Pre-flare filament activations}
\label{ss:pre-flare}

In the AIA data we count 19 episodes with single or multiple ejections of 
bright material from AR~11515 on July~1 before the eruptive flare; some 
of these were observed by EUVI-B as well. Initially, they all originate 
in the region's LS, but from 05~UT onward, the area of newly emerging 
flux (NEF) increasingly becomes the source of the jets. A double jet 
emission from the LS during 03:40--04~UT in both north and south 
directions deserves mentioning, as co-temporal brightenings occur 
in the S-shaped filament in the enhanced network southwest of the 
AR complex, as well as in\B{, and adjacent to,} a filament channel extending 
from \B{the S-shaped filament} to the west. 
The existence of magnetic connections to these quite 
remote areas, which are marked in Figure~\ref{f:activation_AIA} below, 
is thus indicated. Part of the connections are also seen in the PFSS 
extrapolation (Figure~\ref{f:PFSS}).
Later, these remote areas show the rise of material, associated with 
brightenings and dimmings, in close relation to the main eruptive flare.

\begin{figure}
\centerline{
\includegraphics[angle=0,width = \textwidth,clip=]
{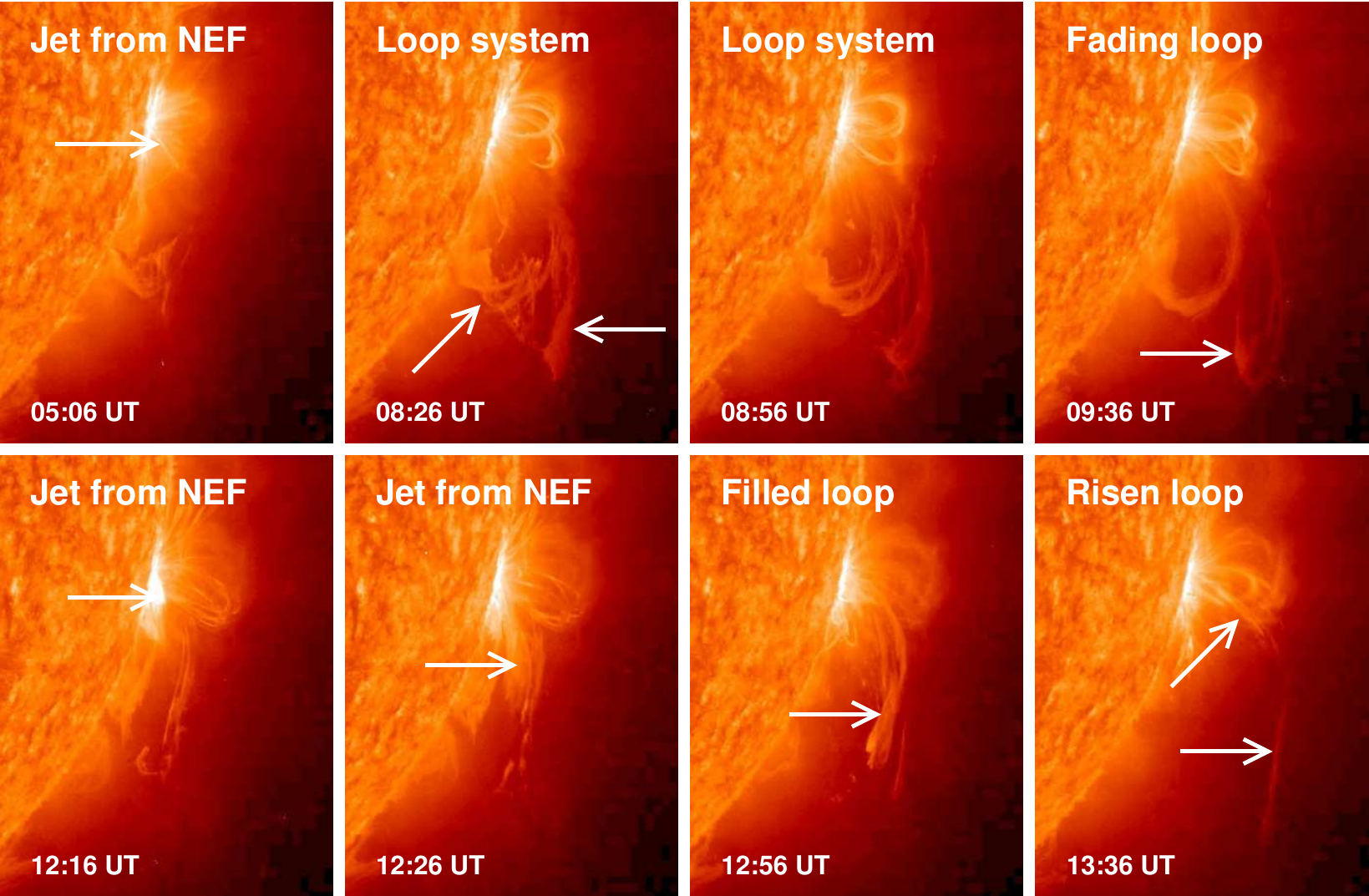}}
\caption{EUVI-B 304~{\AA} snapshots of the \B{two} confined loop and filament 
activations south of AR~11515 during 05--\B{10:30}~UT \B{(top row) and 
12:10--13:40~UT (bottom row), both} before the eruptive 
flare. The rise of \B{three loops south of the AR complex, all} containing 
filament material, is seen \B{following} the ejection of a jet from 
the NEF at $\sim05$ \B{and \B{12:10}}~UT\B{, respectively. The loops fade 
subsequently, mostly due to the draining of the cool material (the lower 
loop seen in panel (d) fades by the time of panels (e)--(f)).}}
\label{f:activation_EUVI}
\end{figure}

The chain of coupled events can be traced back at least to the activation 
of the large loop system containing filament material (dark in 
He~{\sc ii} 304~{\AA}) south of AR~11515, which commenced at $\sim05$~UT, 
i.e. with the first jet ejected from the NEF area. AIA 304 and 193~{\AA} 
images show that the jet is ejected into a large loop connecting the 
\B{area of the} LS and 
the negative polarity of AR~11515 and extending to about S35. The loop 
then expands in the south, east, and west directions, reaching $\sim$~S50 
by 10:30~UT, when it fades in the 304~{\AA} images. A second\B{, somewhat 
smaller loop, lying below the first one in the EUVI-B limb view}
and seen only in the 304~{\AA} band, also starts 
to expand. These loops are relatively faint in the AIA images, where 
they are most clear in animated image sequences, but are clearly seen 
in the limb view of STEREO Behind. The EUVI-B images 
\B{(Figure~\ref{f:activation_EUVI}(a--d))} show the rise of 
a multithreaded loop, filled with He {\sc ii}-emitting plasma and 
apparently rooted in the AR complex and about $10^\circ$ farther 
south, from 05~UT onwards. The threads in the southern part, which 
in part obviously belong to the \B{smaller} loop in the AIA images, show 
\B{indications} 
% clear signs 
of helical structure and undergo various changes until 
the flux is organized in two loops, which reach maximum heights of 
$\sim0.15$ and $\sim0.25~R_\odot$ during \B{9--10}~UT. \B{Both} 
loops then fade at \B{these positions, the upper one by $\approx$~10:20~UT,
the lower one about two hours later}. This 
filament activation appears to be confined. 

\begin{figure}
\centerline{
\includegraphics[angle=0,width = 0.79 \textwidth,clip=]{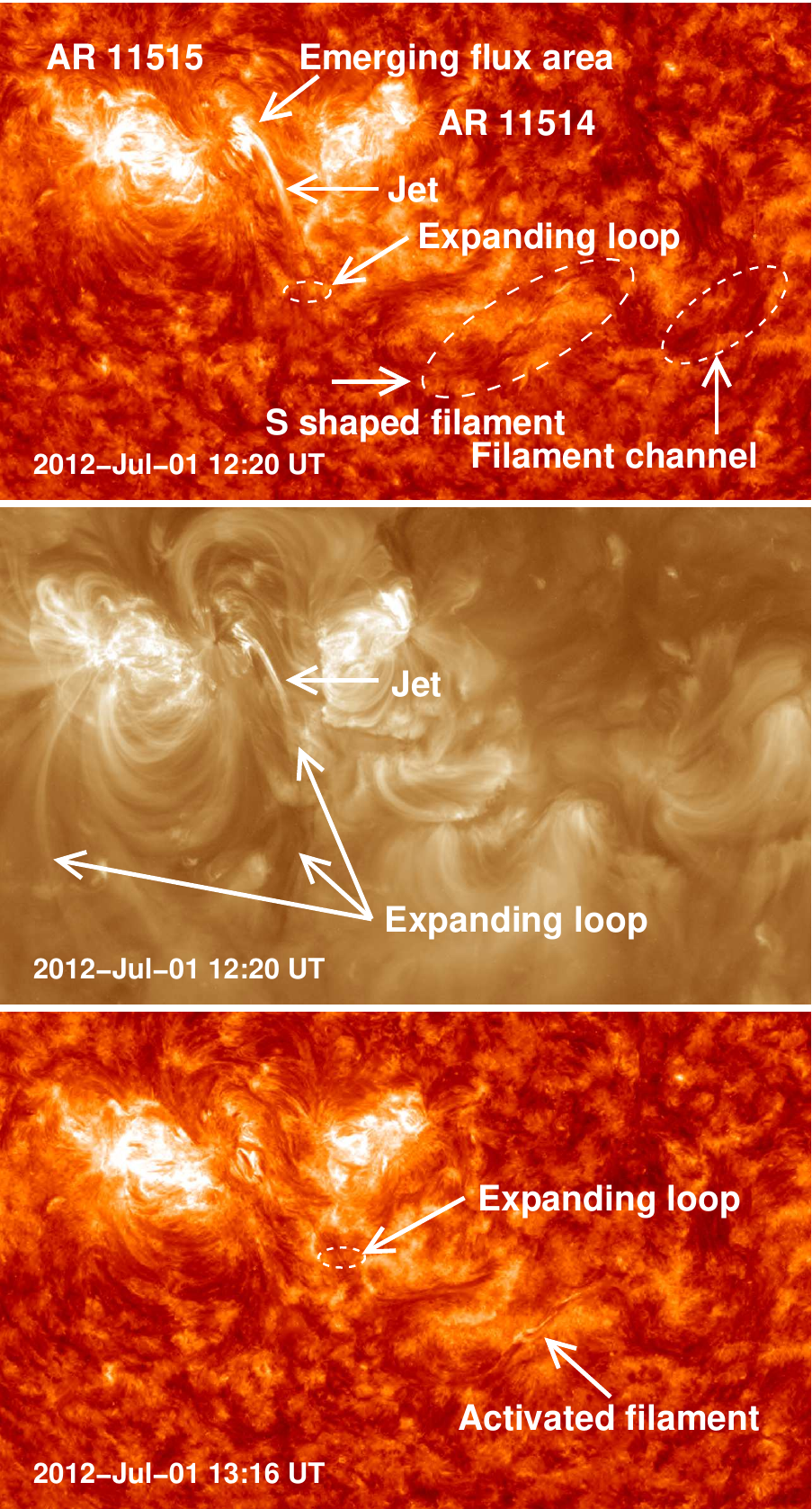}}
\caption{AIA 304 and 193~{\AA} snapshots of the second activation, that 
directly leads to the eruptive flare. Several thin threads of absorbing 
(filament) material in the expanding loop \B{system} are enclosed by 
\B{small} ellipses in the 304~{\AA} images. \B{At 193~{\AA} the legs of 
the expanding loop system are clearly seen in the still image, while the 
top part can be discerned against the structured background only in 
animated image sequences. Two large ellipses outline the positions of the 
S-shaped filament and the filament channel westward of it, which become 
involved in the eruption.}
}
\label{f:activation_AIA}
\end{figure}

Of the further jet emissions in AR~11515, the southward ones from the NEF 
area commencing at about 10:50 and 12:10~UT trigger similar effects in the 
AR complex, which eventually evolve into the eruptive flare. The initially 
slow expansion of a similar southward-oriented loop (lying inside the 
previously expanded one in the AIA images) is triggered by the first jet 
and accelerated by the second. This loop shows a stronger westward 
expansion. Its western leg sweeps over the S-shaped filament in the 
enhanced network southwest of the AR complex, reaching it at 13:12~UT, 
simultaneously with the onset of brightenings in the filament 
(Figure~\ref{f:activation_AIA}). The loop 
fades in the AIA 304~{\AA} images by 13:42~UT, when it crosses over the 
filament roughly in the middle. The jet at 12:10~UT and subsequent loop 
expansion are also imaged by EUVI-B {\color{black}(Figure~\ref{f:activation_EUVI}(e--h))}. 
These 304~{\AA} images show that the 
jet \B{is} eject\B{ed} into the height range of the previously activated 
loop ($\sim0.15~R_\odot$). The He~{\sc ii}-emitting material forms a new 
\B{multi-threaded} loop, which quickly rises 
to $\sim0.4~R_\odot$, but \B{stops at this height. The material then slides 
northward, back to the surface and the loop fades, disappearing at}
$\approx$~14:00~UT; this 
activity apparently also remains confined. Nevertheless, the flux above 
the legs of this loop, which become nearly vertical, from 
$\sim25^\circ$ to the surface at the \B{time of the} jet ejection to $\sim70^\circ$ at the 
fading, must be substantially disturbed. Primarily the western leg, 
running under the middle field lobe of the AR complex, must be seen 
in the EUVI-B 304~{\AA} data, because this leg is much clearer than 
the eastern leg in the corresponding AIA data. 
The loops seen above the AR complex in \B{the EUVI-B 304~{\AA}} images extend 
\B{within the area of} the two ARs, but do not pass over the \B{expanding} loop 
(Figure~\ref{f:activation_AIA}); therefore, they are not lifted. 

The corresponding EUVI-B 195~{\AA} images 
show the 
rise of diffuse, multiple loop structures south of the AR complex 
from $\sim11$~UT, also stalled at $\sim0.4~R_\odot$ by $\sim14$~UT. 
\B{These are not displayed here, since they show the rising structures 
only very faintly.}

The second activation of the loop system south of AR~11515 and its strong 
westward expansion
trigger several brightenings in the S-shaped filament in the enhanced 
network southwest of AR~11514, and fast motions of brightened material 
along the axis of the filament. At least one thread detaches and moves 
away from the filament in a perpendicular (southwest) direction, followed 
by a pair of moderately intense ribbons developing on either side of the 
filament from 13:48~UT onward \B{(these can still be seen 90~min later
in Figure~\ref{f:flare_AIA}(a))}. Clearly, there is an eruption of flux 
in the channel that includes this filament, but is mostly lying above the 
filament. A partial eruption of the S-shaped filament is also obvious from its 
reduced darkness in the H$\alpha$ images. The 
AIA 193~{\AA} images reveal more clearly that the corona above the S-shaped 
filament strongly expands southward as the ribbons to the sides of the 
filament spread, suggestive of the opening of the field in this area, 
which dims subsequently.

\B{Related activities also occur in and adjacent to} the filament channel 
extending westward from the \B{S-shaped} filament. \B{These include} 
a very localized ejection of bright material \B{near its western end} 
during 12:45--13~UT, \B{precisely at the location of a brightening caused 
by the earlier activation during 03:40--04~UT, and} a southward-directed 
eruption \B{from this filament channel} essentially simultaneous with the 
development of the ribbons \B{of the S-shaped filament. This eruption} 
leads to a deep and extended dimming. \B{The locations of} these 
developments \B{in the area of enhanced network are identified in} 
Figures~~\ref{f:activation_AIA} and \ref{f:flare_AIA}.

\subsection{The eruptive flare}
\label{ss:flare}

\begin{figure}
\centerline{
\includegraphics[angle=180,width = 0.7\textwidth]{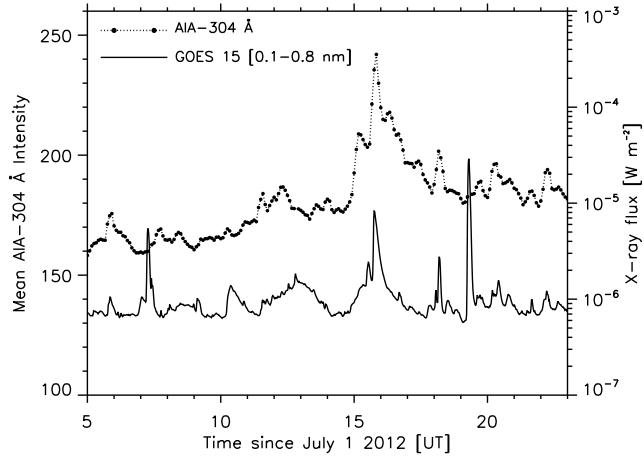}}
\caption{Temporal evolution of the mean AIA 304~\AA~intensity (dotted 
line with filled circles) in the active region and the GOES~15 soft 
X-ray flux in the 0.1--0.8~nm band (solid line). Both the C5.4 
flare at 07:11~UT and the M2.8 flare at 19:11~UT occurred in AR~11513.}
\label{f:GOES}
\end{figure}

\begin{figure}
\centerline{
\includegraphics[angle=0,width = 0.8 \textwidth,clip=]{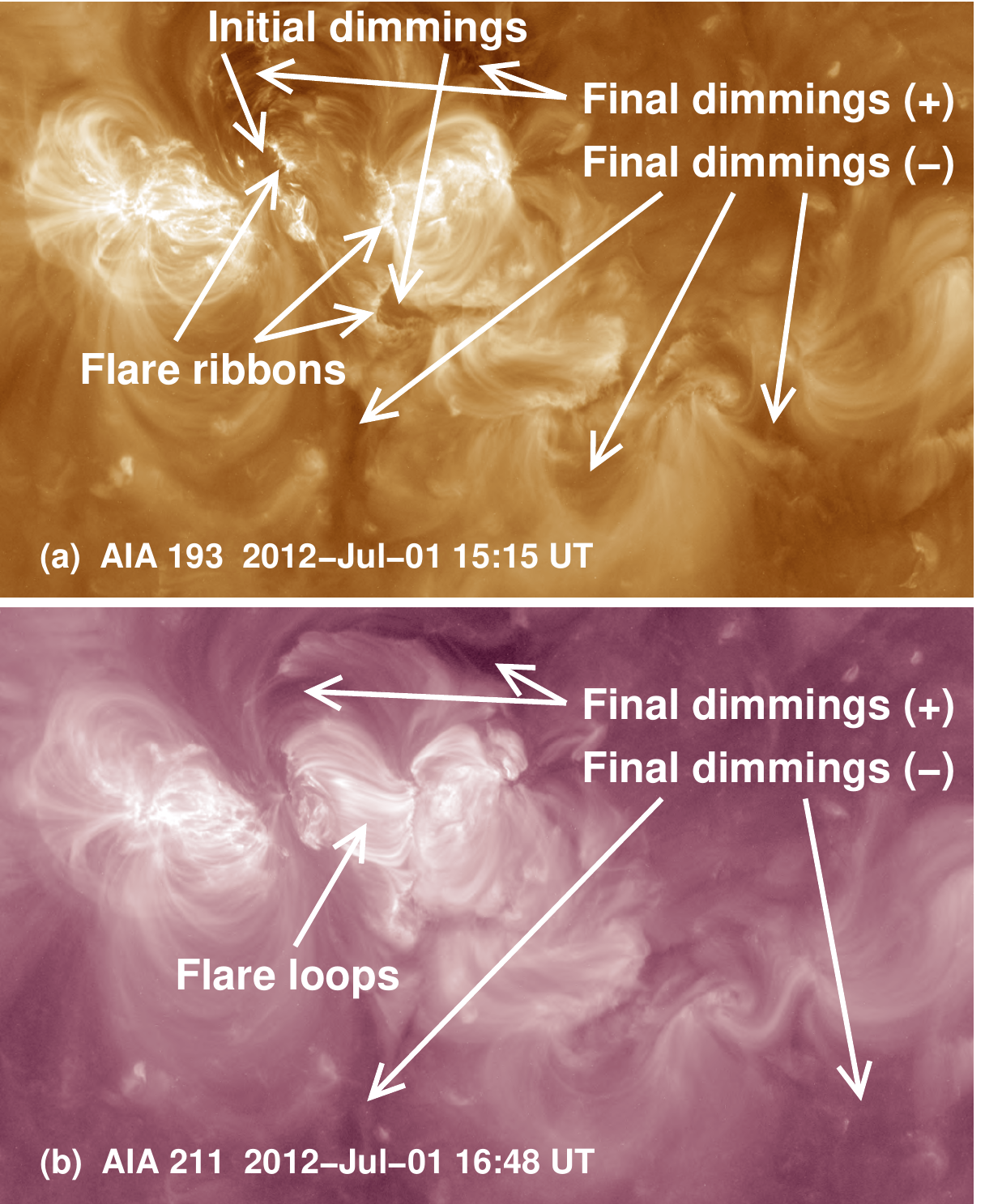}}
\caption{AIA images of the early (top, 193~{\AA}) and late (bottom, 
211~{\AA}) evolution of the eruptive C8.3 flare. The magnetic polarity 
of the dimming areas is indicated.}
\label{f:flare_AIA}
\end{figure}

Although the pre-eruptive and eruptive activity extends over the whole 
AR complex and the large neighboring area of enhanced network to the 
southwest, the core of the eruption clearly is the PIL between 
ARs~11514 and 11515. The bright ribbons begin to form near the 
ends of this PIL, where they enclose the initial dimmings, and 
the ribbons and
flare loop arcade extend along this PIL. Subsequently, extended dimmings, 
interspersed with small-scale brightenings, form in other places, 
indicating that the overlying flux had a complex topology and that 
the erupting flux reconnected with overlying flux, exchanging the 
footpoints.

The ribbons of the eruptive C8.3 flare, which peaks at 15:45~UT 
(Figure~\ref{f:GOES}), begin to form in the positive flux ahead 
and northwest of the LS in AR~11515 at 14:30~UT and about 8 minutes 
later at the \B{eastern} edge of the most \B{eastern} enhanced network cell 
south of AR~11514 \B{(Figure~\ref{f:flare_AIA}(a))}. 
This onset time coincides with the time the ribbons 
at the activated S-shaped filament in the enhanced network have reached 
their maximum separation. Thus, there is a continuous sequence of events, 
starting with jet ejections at the NEF, continuing into the activation 
and rise of a large loop that runs under the middle lobe of the AR 
complex and at the same time sweeping over the area of enhanced 
network. This is followed by the partial eruption of the S-shaped 
filament and the adjacent filament channel in the enhanced network, 
and finally by the destabilization of the flux in the middle lobe of the AR 
complex. The arc-shaped initial section of the ribbon in the positive-flux 
area encloses the initial dimming area, which includes about one-third 
of the LS. On the other side, the initial dimming extends over half of 
the network cell bounded by the initial ribbon. These are the best 
candidates for the footpoint regions of the destabilized flux. They 
are marked in Figure~\ref{f:flare_AIA}(a).

The erupting flux can be discerned only partly and is extremely 
difficult to visualize, since it \B{does not contain} 
any cool material. Animated
images of AIA 193, 211, and 335~{\AA} display a \B{very faint and diffuse} 
loop that propagates westward from the middle field lobe of 
the AR complex, up to $\sim20^\circ$ west from the PIL, during 
14:15--15:15~UT. The loop appears to be rooted in or near the 
initial positive flare ribbon, i.e. in or near the LS. In the southward 
direction it cannot be traced \B{beyond} the edge of AR~11514
\B{against} the complex and changing background. This is consistent with 
the dimming at the initial negative flare ribbon being its southern 
footpoint. The two initial dimmings deepen synchronously with the 
westward motion of the loop. The detection 
in the 193, 211, and 335~{\AA} bands, and the non-detection in the 
94 and 131~{\AA} bands, indicate a temperature in the range 
$1.5~\mbox{MK}\lesssim T<7$~MK. The structure appears to be 
similar to the hot channel detected in other events, mostly 
above the limb \cite[e.g.,][]{2012NatCo...3E.747Z}, with the 
lower temperature here corresponding to the moderate intensity 
of the flare. The hot channel is assumed to show the erupting flux rope.

Within the next half hour, the ribbons quickly extend along the 
PIL, especially the ribbon in the weaker negative flux, which 
then covers the whole north-south extent of negative flux at 
the trailing side of the enhanced network and AR~11514 
\B{(this is indicated in Figure~\ref{f:flare_AIA}(a))}. The 
other ribbon extends into positive network north of the LS 
and propagates toward the center of the LS. Simultaneously, 
several dimming areas begin to develop. These include an 
extended area of positive network north of the unstable 
middle flux lobe (where a prominent set of interconnecting 
loops existed before the eruption and was strongly disturbed), 
the erupting S-shaped filament and adjacent filament channel 
in the enhanced network southwest of AR~11514, and a substantial 
area of negative network extending south from the initial 
negative ribbon \B{(Figure~\ref{f:flare_AIA}(b))}. 
These dimming areas last for several hours. 
The three southern ones formed in the first place during the 
pre-eruption activities (rise of the large loops south of 
AR~11515 and rise of flux above the S-shaped filament and 
adjacent filament channel southwest of AR~11514 \B{from 13:12~UT onward}), but extend 
farther in the course of the eruptive flare. They indicate that 
the erupting flux reconnect\B{s} and exchange\B{s} footpoints with 
originally overlying flux rooted in at least some of the dimming 
areas.

\begin{figure}
\centerline{
\includegraphics[angle=0,width = 0.6 \textwidth,clip=]
{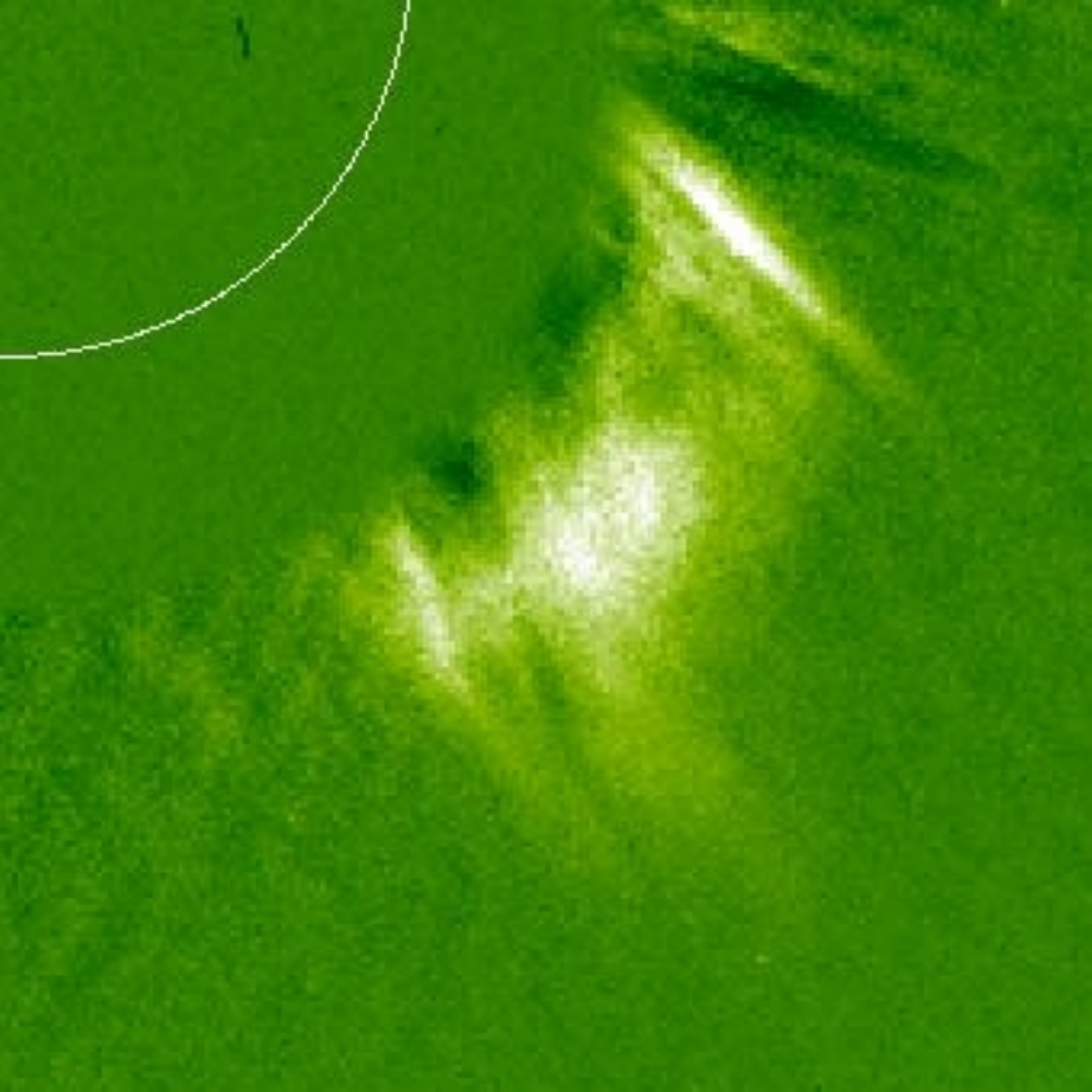}}
\caption{
COR1-B running-difference image at 15:30~UT showing the 
considerable southward angular extent of the erupting flux, consistent 
with it being rooted in the negative-polarity dimming areas south of 
the AR complex, which are marked in Figure~\ref{f:flare_AIA}.}
\label{f:CME}
\end{figure}

This is supported by the EUVI-B 195~{\AA} images of the eruption, 
which show a loop expanding 
radially and southward above the AR complex from about 14:40~UT onward 
(then at a height of $\sim0.3~R_\odot$). The loop appears to be rooted 
in the AR complex and up to $20^\circ$ south of it (the precise location 
being masked by foreground and background loops). This is consistent 
with the dimmings in the positive network regions southwest and south 
of the AR complex. \B{Since} the loop in the EUVI-B
195~{\AA} images is too faint for a visualization in still images,
even when using difference images, Figure~\ref{f:CME} shows a COR1-B
difference image of the CME at an early stage, when it still expands
in the same latitude range as the loop imaged by EUVI-B at 195~{\AA}.

The direction of the flux overlying the middle field lobe of the AR 
complex, implied by the location of the dimmings in AIA images, 
is \B{roughly from north-east to south-west, \i.e.} 
rather close to the direction of the flux in the lobe, especially 
if the sheared loops in the lobe are considered. This is different 
from the configuration of the breakout model for eruptions.

The arcade of flare loops forms at the section of the PIL between the 
two ARs, supporting the view that the destabilization of sheared flux 
in the middle lobe of the AR complex \B{caused} 
the eruption. The usual evolution from considerably sheared first flare 
loops to more potential ones later on can be seen 
\B{(Figures~\ref{f:flare_AIA}(b) and \ref{f:add-flare}(b--d))}. The first flare 
loops form after 14:38~UT in AIA 94 and 131~{\AA} images 
($T\sim7\mbox{--}11$~MK).

A fast, southward-propagating CME appears in the LASCO/C2 field of 
view at 15:35~UT. A second-order fit yields a velocity of 
800~km\,s$^{-1}$ at the heliocentric distance of $3~R_\odot$. The 
propagation direction is south\B{west}ward in the field of view of 
COR1-B (Figure~\ref{f:CME}), where the CME appears at 15:00~UT.

The C2.4 flare peaking at 15:32~UT, i.e. during the rise of the 
flare in the AR complex studied here, is a compact and confined, 
likely sympathetic flare in AR~11513.

\subsection{Superimposed confined flaring activity}
\label{ss:add-flare}

\begin{figure}
\centerline{
\includegraphics[angle=0,width = 0.6 \textwidth,clip=]{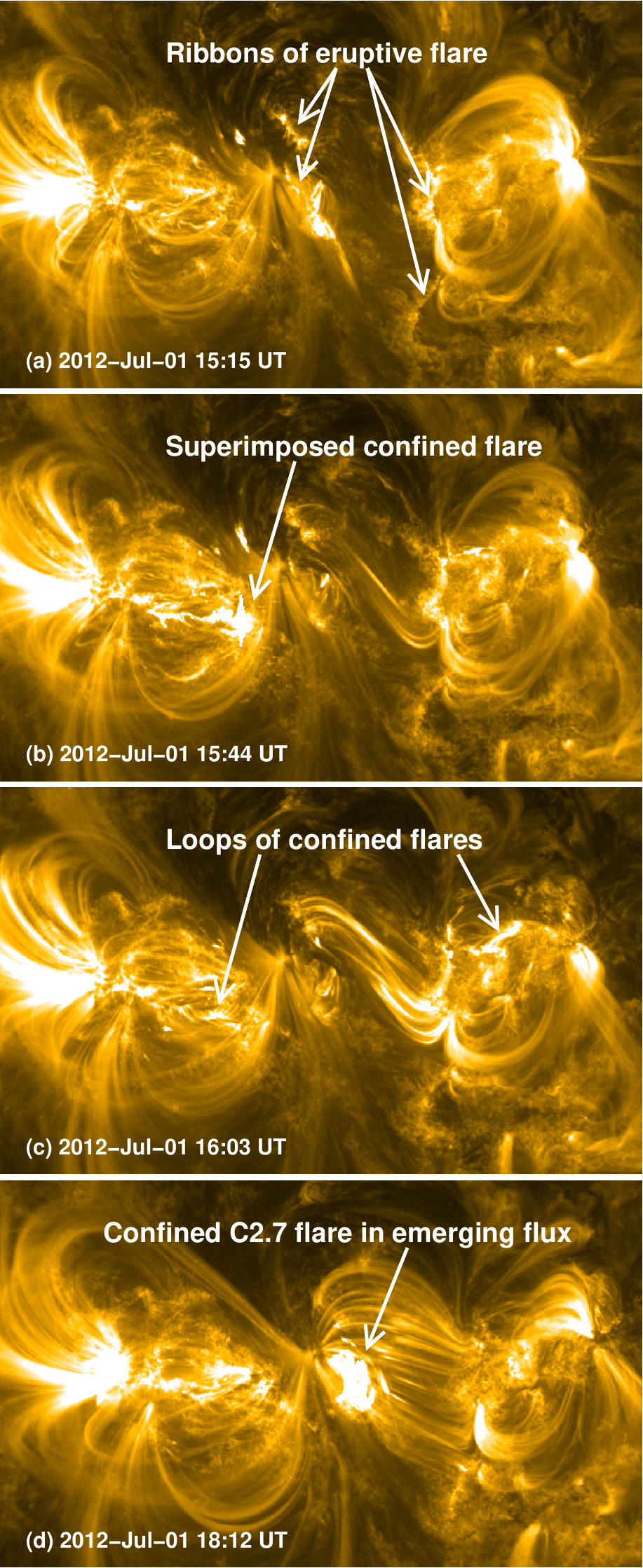}}
\caption{AIA 171~{\AA} images of the additional confined flares. 
(a) Early phase of the eruptive flare, before the confined flares; 
(b) peak emission of the combined eruptive and first additional confined 
flares; (c) loops of the two confined flares in AR~11515 and AR~11514 
after their peak times; (d) peak emission of the confined C2.7 flare 
in the NEF under the arcade of the eruptive flare.}
\label{f:add-flare}
\end{figure}

\B{Two} confined flares, one in AR~11515, the other in AR~11514, 
are closely related to the eruptive flare. Both occur under the closed, 
main field lobe of their bipolar host AR. The one in AR~11515 marks 
the peak of the emission of the C8.3 flare at 15:45~UT, i.e., is 
brighter than the eruptive flare. However, the confined flare 
is clearly a separate event, occurring in a different lobe of flux, 
where it remains compact and confined, and \B{developing} on a much shorter 
timescale. It is located at the PIL formed by the westward tongue of 
negative flux in the middle of the AR.
In the AIA 131~{\AA} band, it develops only from 15:37~UT onward, one 
hour after the onset of the eruptive flare. The other confined flare 
occurs within 5--10~minutes from the first at the northern end of the 
internal PIL in AR~11514. These confined flares are most sharply imaged 
in the 171~{\AA} band, as shown in Figure~\ref{f:add-flare}; they are 
of course more prominent, but more diffuse, in the ``hot'' AIA 
bands, up to 131~{\AA}. Both may be triggered by changes in their 
overlying flux caused by the rise of the flux in the eruptive flare. 
In particular, material is seen rising from a position 
very near the second flare immediately before its onset.

A third confined flare, of X-ray class C2.7, occurs in the NEF 
during the decay of the eruptive flare, i.e. under the flare loop 
arcade (see also Figure~\ref{f:add-flare}). Its loops are oriented 
north-south, as is \B{part of} the field rooted in the newly emerged flux ahead 
of the LS (Section~\ref{s:source}). A detailed inspection of the 
171~{\AA} data suggests that the NEF 
brightens due to reconnection with loops rooted under the flare 
loop arcade, which show brief brightenings and changes in shape 
(see the 171~{\AA} data in the animation accompanying 
Figures~\ref{f:overview1} and \ref{f:overview2}.)
Most likely, the continuously rising NEF pushes against the 
re-closed field of the middle lobe in the AR complex, and the 
temporal association with the decay phase of the eruptive flare 
is accidental.

\section{Discussion: initiation of the eruption}
\label{s:discussion}

The eruption occurs at the PIL between ARs~11514 and 11515, under 
the middle lobe of a three-lobe magnetic configuration, where the field 
is considerably sheared and a filament channel is forming. Moreover, 
new flux is emerging under the lobe near its edge. The amount of 
new flux emerged by the time of the eruption is only 
$\sim40\,\%$ of the flux in the middle lobe. The 
observations are inconclusive with regard to the question whether 
a flux rope was formed (or is forming) above the PIL, or whether 
the flux has the structure of a sheared arcade.

Both basic possibilities for the triggering of eruptions appear 
consistent with the data: the equilibrium of sheared 
(current-carrying) flux can be destroyed either by an increase 
of the current beyond the limit of stability, or by a weakening 
of the external \B{flux} (rooted in photospheric sources other 
than the sheared flux) by a sufficient amount, such that \B{it can no 
longer stabilize} the upward-directed Lorentz self-force of the given 
current. The first option is supported by the flow 
toward the PIL from the \B{trailing} side of AR~11514. This transports 
negative flux to the PIL, some of which cancels there with 
positive network flux, increasing the shear and current above 
the PIL. However, by the amount of flux involved 
{\color{black}(around $1.5\times 10^{18}$~Mx\B{, $\approx450$ times 
smaller than the amount of emerging flux})}, 
this is a very weak effect compared to the multitude of other pre-flare 
activities that affect the large-scale field structure of 
the AR complex. Therefore, we consider such ``internal triggering'' 
at the PIL to be \emph{far} less likely than an external triggering 
by changes of the ambient field.

A classical model for external triggering in a quadrupolar source 
region is the breakout model \citep{2008ApJ...683.1192L, 2012ApJ...760...81K}. 
This model relies on the weakening of the ambient flux in the middle lobe by 
reconnection with an overlying fourth lobe of essentially opposite direction. 
We do not find signatures that such a lobe exists and contains a sufficient 
amount of flux. The PFSS extrapolation is inconclusive with regard to the 
existence and direction of such a field lobe, but indicates that not much 
flux could be contained if it exists. The dimmings developing in association 
with the eruption indicate that the overlying flux is directed relative 
to the middle field lobe at an angle smaller than $90^\circ$, basically 
opposite to the model. Therefore, it is unlikely that the breakout model 
can explain the onset of the eruption.

The event yields many strong indications that the episode of newly emerging 
flux under the middle flux lobe triggers interactions with the large-scale 
field in the AR complex, which lead to an upward expansion of the middle 
lobe, weakening its stabilizing downward tension force. The most likely scenario 
in our opinion is that the rise of the large loop system south of AR~11515, 
which is rooted near the area of NEF and the LS and initially runs at a low 
elevation, i.e. essentially under the middle flux lobe, lifts the lobe quite 
strongly (see, in particular, {\color{black}Figure~\ref{f:activation_EUVI}(f--h))}. 
Additionally, the expansion 
of the loop system probably triggers the eruption of the S-shaped filament in 
the enhanced network southwest of AR~11514, and of the filament channel 
extending westward from the filament. These secondary eruptions perturb 
the large-scale field structure in the AR complex, as is obvious from the 
development of dimmings, and very likely have a share in lifting the middle 
flux lobe. This is indicated by their close temporal association with the 
main eruption, which commences while they are in progress. Each of the 
relevant pre-eruption activities commences with, or is amplified by, a 
jet emission from the NEF. The jets are emitted \emph{into} the large 
loop system south of AR~11515, which thus has become magnetically 
connected with the NEF through reconnection. The loop system must 
carry currents already before the eruption, since the PFSS extrapolation 
does not yield any low-lying field lines at the location of the loop 
system. There is no doubt that the currents, and consequently the 
equilibrium of the loop system, are strongly perturbed when the loop 
system reconnects with the NEF.

Thus, we trace the origin of the eruption to the NEF, one of the 
standard options for eruption triggering \cite[e.g.,][]{2009AdSpR..43..739S}. 
The particular trigger mechanism in the present event, however, 
is partly different from the mechanisms considered so far. Both 
observations \citep{1995JGR...100.3355F} and modeling find the 
triggering of eruptions by emerging flux to be most efficient 
when the new flux is directed oppositely to the existing flux, 
such that reconnection commences easily. This is true for  
emergence both near the PIL and near the edge of the arcade field 
overlying the PIL, although the latter case was found to be 
restricted in the range of distances to the PIL \citep{2008ChA&A..32...56X}. 
The reconnection destabilizes a pre-existing flux rope 
\citep{2000ApJ...545..524C, 2005ChJAA...5..636X} or transforms the inner part of 
a pre-existing sheared arcade into an unstable flux rope 
\citep{2012ApJ...760...31K, 2014ApJ...796...44K}. Alternatively, 
the emerging flux itself forms a flux rope, and the reconnection 
of this flux rope with a pre-existing arcade is essential for 
the removal of the rope's own stabilizing envelope, such that 
the  rope will erupt 
\citep{2008A&A...492L..35A, 2012A&A...537A..62A, 2014ApJ...787...46L}. 
None of these mechanisms appears to be at work in the eruption 
studied here. The dominant orientation of the NEF is \B{between westward 
and southward}, not much different from the orientation of 
the pre-existing flux in the middle field lobe, so that the 
mechanisms based on the occurrence of reconnection \B{with 
antiparallel or nearly antiparallel field}, elaborated 
in the above articles, are not applicable. Additionally, the amount 
of flux emerged by the time of the eruption is considerably
smaller than the pre-existing flux, clearly insufficient 
for the \B{alternative} arcade compression mechanism studied by 
\citet{2014ApJ...796...44K}. 

Our suggestion of destabilization by lifting the overlying flux 
bears some similarity to the findings in \citet{1999ApJ...510L.157W}. 
These authors used a superposition of potential (vacuum) fields to 
model the effect of new flux emergence and obtained a lifting or 
sidewards deflection of flux overlying filaments that erupted. 
This is conceptually different from the lifting of the middle 
flux lobe by the rise of the activated loop \B{system} in our suggested 
scenario, which must be driven by Lorentz forces, but has a 
similar effect on the stability of the sheared flux in the 
filament channel.

The NEF in the present event does reconnect with overlying, 
pre-existing flux, but this flux is not part of the later 
erupting middle lobe of the quadrupolar AR complex, but rather 
part of a side lobe (AR~11515). This flux, seen as a large loop 
system south of AR~11515, and rooted at both ends in this AR, 
runs under one side of the middle lobe. Its direction is not much 
different from the field direction in the middle lobe, and no signs 
of reconnection with the middle lobe (any brightenings other than 
the jet ejection from the NEF into the loop system) are seen in 
either AIA or EUVI-B images. \B{This indicates that the rise of the 
loop system} and the consequent 
lifting of the flux in the middle lobe is the dominant effect 
that destabilizes the lobe.

Although the magnetic topology of the eruption may be a relatively 
rare occurrence, it need not be a singular case. Configurations 
with both interconnecting loops to a preceding active region and 
internal loops in an active region emerging from the leading 
sunspot are common. New flux is often seen to emerge 
in the vicinity of existing flux concentrations.

The observations do not yield a conclusive hint on whether or 
not a flux rope already exists by the time the middle lobe is 
lifted. If none existed before, a rope must form in response 
to the lifting, because otherwise there would be no 
eruption---there is no ideal MHD instability of a sheared 
arcade \cite[e.g.,][]{1994ApJ...423..847R, 
1994ApJ...430..898M, 1996A&A...306..913A}. One can expect 
that the lifting of a sufficiently sheared arcade by 
external forces induces Lorentz forces in the lifted 
flux that point toward the middle of the arcade and 
lead to a horizontal constriction that steepens a 
vertical current sheet in the center of the arcade. 
This is similar to the enforced rise of an arcade over 
an unstable flux rope, which leads to a constriction of 
the flux under the rope, current sheet steepening, and 
eventually to flare reconnection. Here the reconnection 
would form a flux rope if none existed before and then 
continue in the standard manner of flare reconnection, 
amplified by the eruption of the formed rope.

\B{Following perturbations by the NEF, part of the loop system running
south of AR~11515 and under the middle lobe begins to rise. However, 
all loops find a new equilibrium height (Figure~\ref{f:activation_EUVI}),
whereas the sheared flux under the middle lobe erupts into a CME. This
difference is likely related to a difference in the nature of the two
structure's force-free equilibria. We suggest the following
explanation. The current in the sheared flux rooted near the PIL is
likely stabilized by the overlying flux of the middle lobe rooted to a
large part away from the PIL at the edges of the adjacent ARs. This is
the classic force-free equilibrium of flux carrying a net current in
external, essentially potential flux, as proposed by
\citet{1978SoPh...59..115V} and elaborated by
\citet{Titov&Demoulin1999}. This equilibrium admits an instability
(torus instability) when the sheared flux transforms, at least partly,
into a flux rope above the PIL \citep{1978SoPh...59..115V,
2006PhRvL..96y5002K}. It is difficult to see how such an equilibrium
could be established for the loop system south of AR~11515, because
there is hardly any significant amount of flux rooted in the area
south of the loop system (Figure~\ref{f:overview2}). Therefore, the
equilibrium of the loop system, which also carries electric currents
(as argued above), is likely provided by a return current such that
the net current along the loop system vanishes. Such force-free
equilibria exist for arcades and flux ropes; the flux rope can have
the shape of a loop. While the arcades are stable, flux ropes may
admit an instability of the helical kink mode. However, the existence
of instability for such flux ropes is not yet clarified
\citep{Torok&Kliem2003, 2005A&A...430.1067A}. If instability
occurs, a minimum twist of about one field line turn is required. The
EUV images of the loop system do not provide any evidence supporting
this substantial requirement (see Figures~\ref{f:overview2} and 
\ref{f:activation_AIA}). Therefore, the loop system is probably in a
very stable force-free equilibrium. Its reconnection with the NEF must
change its currents, such that it acquires a new equilibrium at a
higher position without erupting.}

\section{Conclusions}
\label{s:conclusions}

The present investigation of an eruptive C8.3 flare and fast CME 
on July~1, 2012 in the AR complex consisting of ARs~11514 and 
11515 reveals a new scenario of eruption triggered by emerging 
flux. The emergence occurs under the middle flux lobe of a 
quadrupolar configuration (between the two ARs) far away from the 
PIL. Most of the emerging flux is roughly aligned with the 
\B{arcade-like large-scale flux in this lobe}, not favorable for reconnection 
with it. Moreover, the amount of emerging flux remains 
moderate, reaching only $\sim40\,\%$ of the 
pre-existing flux of the lobe by the time of the eruption. A 
relatively particular circumstance is given by the fact that 
a large loop system of the trailing AR~11515, rooted in the 
leading spot of the region, passes under the middle lobe, which 
is also largely rooted in this spot. The new flux emerges in 
front of the spot, under the loop system. It reconnects with 
the loop system, ejecting a series of jets into the loop 
system, which \B{then} rises considerably (changing its inclination 
to the surface from $25^\circ$ to $70^\circ$) but does not 
erupt by itself. The rise of the loop system lifts the flux 
of the middle lobe, reducing its stabilizing downward-tension. 
This triggers the eruption of the \B{strongly} sheared flux in the 
center of the lobe.

The eruption is also noteworthy for its location. The structure 
of the magnetogram would hardly suggest that an eruption occurs 
in the extended area of very weak field between the two ARs, 
one of them already decaying. Instead, one would expect an 
eruption to occur in the complex field of the developing AR, which 
also harbors the emergence of new flux in the immediate vicinity 
of its leading spot (where an eruption, analyzed in {\my}, indeed 
occurs 19~hours later). However, the chromospheric and coronal 
structures indicate magnetic shear at the PIL between the ARs, 
correctly pointing at the site of the eruption studied here.

\begin{acks}
\B{We gratefully acknowledge constructive comments by the referee, which 
were helpful in improving the clarity of this paper.}
R.E.L.\ is grateful for
the financial assistance from the German Science Foundation (DFG)
under grant DE 787/3-1 and the European Commission's FP7 Capacities
Programme under the Grant Agreement number 312495.
G.C.\ and B.K.\ acknowledge support by the NSF under Grant 
No.\ 1249270. B.K.\ also acknowledges support by the DFG.
HMI data are courtesy
of NASA/SDO and the HMI science team. They are provided by the Joint
Science Operations Center -- Science Data Processing at Stanford
University. 
\B{EUVI-B and COR1-B images are supplied courtesy of the STEREO Sun Earth 
Connection Coronal and Heliospheric Investigation (SECCHI) team.}
This work utilizes data obtained by the Global Oscillation
Network Group (GONG) Program, managed by the National Solar
Observatory, which is operated by AURA, Inc.\ under a cooperative
agreement with the National Science Foundation. The data were
acquired by instruments operated by the Big Bear Solar Observatory,
High Altitude Observatory, Learmonth Solar Observatory,
Udaipur Solar Observatory, Instituto de Astrof\'isica de Canarias,
and Cerro Tololo Interamerican Observatory.
We have used the SOHO/LASCO CME catalog, generated and maintained at the
CDAW Data Center by NASA and The Catholic University of America in
cooperation with the Naval Research Laboratory. SOHO is a project of
international cooperation between ESA and NASA.

\end{acks}

\bibliographystyle{spr-mp-sola}
\bibliography{louis_reference_v6BK}

\begin{thebibliography}{61}
% BibTex style file: spr-mp-sola.bst (nameyear), 2014-02-13
\ifx\bisbn     \undefined \def\bisbn  #1{ISBN #1}\fi
\ifx\binits    \undefined \def\binits#1{#1}\fi
\ifx\bauthor   \undefined \def\bauthor#1{#1}\fi
\ifx\batitle   \undefined \def\batitle#1{#1}\fi
\ifx\bjtitle   \undefined \def\bjtitle#1{\textit{#1}}\fi
\ifx\bvolume   \undefined \def\bvolume#1{\textbf{#1}}\fi
\ifx\byear     \undefined \def\byear#1{#1}\fi
\ifx\bissue    \undefined \def\bissue#1{#1}\fi
\ifx\bfpage    \undefined \def\bfpage#1{#1}\fi
\ifx\blpage    \undefined \def\blpage #1{#1}\fi
\ifx\burl      \undefined \def\burl#1{\textsf{#1}}\fi
\ifx\href      \undefined \def\href#1#2{\textsf{#2}}\fi
\ifx\betal     \undefined \def\betal{\textit{et al.}}\fi
\ifx\bctitle   \undefined \def\bctitle#1{#1}\fi
\ifx\beditor   \undefined \def\beditor#1{#1}\fi
\ifx\bbtitle   \undefined \def\bbtitle#1{\textit{#1}}\fi
\ifx\bedition  \undefined \def\bedition#1{#1}\fi
\ifx\bseriesno \undefined \def\bseriesno#1{\textbf{#1}}\fi
\ifx\blocation \undefined \def\blocation#1{#1}\fi
\ifx\bsertitle \undefined \def\bsertitle#1{\textit{#1}}\fi
\ifx\bsnm      \undefined \def\bsnm#1{#1}\fi
\ifx\bsuffix   \undefined \def\bsuffix#1{#1}\fi
\ifx\bparticle \undefined \def\bparticle#1{#1}\fi
\ifx\barticle  \undefined \def\barticle#1{}\fi
\ifx\binstitute  \undefined \def\binstitute#1{#1}\fi
\ifx\bpublisher  \undefined \def\bpublisher#1{#1}\fi
\ifx\doiurl    \undefined
  \def\doiurl#1{\href{http://dx.doi.org/#1}{\textsf{DOI}}}\fi
\ifx\arxivurl  \undefined
  \def\arxivurl#1{\href{http://arxiv.org/abs/#1}{\textsf{arXiv}}}\fi
\ifx\adsurl    \undefined
  \def\adsurl#1{\href{http://adsabs.harvard.edu/abs/#1}{\textsf{ADS}}}\fi
\ifx\botherref \undefined \def\botherref#1{}\fi
\ifx\url       \undefined \def\url#1{\textsf{#1}}\fi
\ifx\bchapter  \undefined \def\bchapter#1{}\fi
\ifx\bbook     \undefined \def\bbook#1{}\fi
\ifx\bcomment  \undefined \def\bcomment#1{#1}\fi
\ifx\oauthor   \undefined \def\oauthor#1{#1}\fi
\ifx\citeauthoryear \undefined\def \citeauthoryear#1{#1}\fi
\def\endbibitem {}
\ifx\bconflocation  \undefined \def\bconflocation#1{#1} \fi

\bibitem[\protect\citeauthoryear{{Amari}
  \textit{et~al.}}{1996}]{1996A&A...306..913A}
\begin{barticle}
\bauthor{\bsnm{{Amari}}, \binits{T.}},
\bauthor{\bsnm{{Luciani}}, \binits{J.F.}},
\bauthor{\bsnm{{Aly}}, \binits{J.J.}},
\bauthor{\bsnm{{Tagger}}, \binits{M.}}:
\byear{1996},
\batitle{{Plasmoid formation in a single sheared arcade and application to
  coronal mass ejections.}}
\bjtitle{\aap}
\bvolume{306},
\bfpage{913}.
\adsurl{http://ads.ari.uni-heidelberg.de/abs/1996A\%26A...306..913A}.
\end{barticle}
\endbibitem

\bibitem[\protect\citeauthoryear{{Amari}
  \textit{et~al.}}{2003}]{2003ApJ...595.1231A}
\begin{barticle}
\bauthor{\bsnm{{Amari}}, \binits{T.}},
\bauthor{\bsnm{{Luciani}}, \binits{J.F.}},
\bauthor{\bsnm{{Aly}}, \binits{J.J.}},
\bauthor{\bsnm{{Mikic}}, \binits{Z.}},
\bauthor{\bsnm{{Linker}}, \binits{J.}}:
\byear{2003},
\batitle{{Coronal Mass Ejection: Initiation, Magnetic Helicity, and Flux Ropes.
  II. Turbulent Diffusion-driven Evolution}}.
\bjtitle{\apj}
\bvolume{595},
\bfpage{1231}.
\doiurl{10.1086/377444}.
\adsurl{2003ApJ...595.1231A}.
\end{barticle}
\endbibitem

\bibitem[\protect\citeauthoryear{{Amari}
  \textit{et~al.}}{2011}]{2011ApJ...742L..27A}
\begin{barticle}
\bauthor{\bsnm{{Amari}}, \binits{T.}},
\bauthor{\bsnm{{Aly}}, \binits{J.-J.}},
\bauthor{\bsnm{{Luciani}}, \binits{J.-F.}},
\bauthor{\bsnm{{Mikic}}, \binits{Z.}},
\bauthor{\bsnm{{Linker}}, \binits{J.}}:
\byear{2011},
\batitle{{Coronal Mass Ejection Initiation by Converging Photospheric Flows:
  Toward a Realistic Model}}.
\bjtitle{\apjl}
\bvolume{742},
\bfpage{L27}.
\doiurl{10.1088/2041-8205/742/2/L27}.
\adsurl{2011ApJ...742L..27A}.
\end{barticle}
\endbibitem

\bibitem[\protect\citeauthoryear{{Archontis} and
  {Hood}}{2010}]{2010A&A...514A..56A}
\begin{barticle}
\bauthor{\bsnm{{Archontis}}, \binits{V.}},
\bauthor{\bsnm{{Hood}}, \binits{A.W.}}:
\byear{2010},
\batitle{{Flux emergence and coronal eruption}}.
\bjtitle{\aap}
\bvolume{514},
\bfpage{A56}.
\doiurl{10.1051/0004-6361/200913502}.
\adsurl{2010A\%26A...514A..56A}.
\end{barticle}
\endbibitem

\bibitem[\protect\citeauthoryear{{Archontis} and
  {Hood}}{2012}]{2012A&A...537A..62A}
\begin{barticle}
\bauthor{\bsnm{{Archontis}}, \binits{V.}},
\bauthor{\bsnm{{Hood}}, \binits{A.W.}}:
\byear{2012},
\batitle{{Magnetic flux emergence: a precursor of solar plasma expulsion}}.
\bjtitle{\aap}
\bvolume{537},
\bfpage{A62}.
\doiurl{10.1051/0004-6361/201116956}.
\adsurl{http://ads.ari.uni-heidelberg.de/abs/2012A\%26A...537A..62A}.
\end{barticle}
\endbibitem

\bibitem[\protect\citeauthoryear{{Archontis} and
  {T{\"o}r{\"o}k}}{2008}]{2008A&A...492L..35A}
\begin{barticle}
\bauthor{\bsnm{{Archontis}}, \binits{V.}},
\bauthor{\bsnm{{T{\"o}r{\"o}k}}, \binits{T.}}:
\byear{2008},
\batitle{{Eruption of magnetic flux ropes during flux emergence}}.
\bjtitle{\aap}
\bvolume{492},
\bfpage{L35}.
\doiurl{10.1051/0004-6361:200811131}.
\adsurl{http://esoads.eso.org/abs/2008A\%26A...492L..35A}.
\end{barticle}
\endbibitem

\bibitem[\protect\citeauthoryear{{Aulanier}, {D{\'e}moulin}, and
  {Grappin}}{2005}]{2005A&A...430.1067A}
\begin{barticle}
\bauthor{\bsnm{{Aulanier}}, \binits{G.}},
\bauthor{\bsnm{{D{\'e}moulin}}, \binits{P.}},
\bauthor{\bsnm{{Grappin}}, \binits{R.}}:
\byear{2005},
\batitle{{Equilibrium and observational properties of line-tied twisted flux
  tubes}}.
\bjtitle{\aap}
\bvolume{430},
\bfpage{1067}.
\doiurl{10.1051/0004-6361:20041519}.
\adsurl{http://ads.ari.uni-heidelberg.de/abs/2005A\%26A...430.1067A}.
\end{barticle}
\endbibitem

\bibitem[\protect\citeauthoryear{{Aulanier}
  \textit{et~al.}}{2010}]{2010ApJ...708..314A}
\begin{barticle}
\bauthor{\bsnm{{Aulanier}}, \binits{G.}},
\bauthor{\bsnm{{T{\"o}r{\"o}k}}, \binits{T.}},
\bauthor{\bsnm{{D{\'e}moulin}}, \binits{P.}},
\bauthor{\bsnm{{DeLuca}}, \binits{E.E.}}:
\byear{2010},
\batitle{{Formation of Torus-Unstable Flux Ropes and Electric Currents in
  Erupting Sigmoids}}.
\bjtitle{\apj}
\bvolume{708},
\bfpage{314}.
\doiurl{10.1088/0004-637X/708/1/314}.
\adsurl{2010ApJ...708..314A}.
\end{barticle}
\endbibitem

\bibitem[\protect\citeauthoryear{{Burtseva} and
  {Petrie}}{2013}]{2013SoPh..283..429B}
\begin{barticle}
\bauthor{\bsnm{{Burtseva}}, \binits{O.}},
\bauthor{\bsnm{{Petrie}}, \binits{G.}}:
\byear{2013},
\batitle{{Magnetic Flux Changes and Cancellation Associated with X-Class and
  M-Class Flares}}.
\bjtitle{\solphys}
\bvolume{283},
\bfpage{429}.
\doiurl{10.1007/s11207-013-0241-8}.
\adsurl{2013SoPh..283..429B}.
\end{barticle}
\endbibitem

\bibitem[\protect\citeauthoryear{{Chen} and
  {Shibata}}{2000}]{2000ApJ...545..524C}
\begin{barticle}
\bauthor{\bsnm{{Chen}}, \binits{P.F.}},
\bauthor{\bsnm{{Shibata}}, \binits{K.}}:
\byear{2000},
\batitle{{An Emerging Flux Trigger Mechanism for Coronal Mass Ejections}}.
\bjtitle{\apj}
\bvolume{545},
\bfpage{524}.
\doiurl{10.1086/317803}.
\adsurl{2000ApJ...545..524C}.
\end{barticle}
\endbibitem

\bibitem[\protect\citeauthoryear{{Demoulin}
  \textit{et~al.}}{1993}]{1993A&A...271..292D}
\begin{barticle}
\bauthor{\bsnm{{Demoulin}}, \binits{P.}},
\bauthor{\bsnm{{van Driel-Gesztelyi}}, \binits{L.}},
\bauthor{\bsnm{{Schmieder}}, \binits{B.}},
\bauthor{\bsnm{{Hemoux}}, \binits{J.C.}},
\bauthor{\bsnm{{Csepura}}, \binits{G.}},
\bauthor{\bsnm{{Hagyard}}, \binits{M.J.}}:
\byear{1993},
\batitle{{Evidence for magnetic reconnection in solar flares}}.
\bjtitle{\aap}
\bvolume{271},
\bfpage{292}.
\adsurl{1993A\%26A...271..292D}.
\end{barticle}
\endbibitem

\bibitem[\protect\citeauthoryear{{Feynman} and
  {Martin}}{1995}]{1995JGR...100.3355F}
\begin{barticle}
\bauthor{\bsnm{{Feynman}}, \binits{J.}},
\bauthor{\bsnm{{Martin}}, \binits{S.F.}}:
\byear{1995},
\batitle{{The initiation of coronal mass ejections by newly emerging magnetic
  flux}}.
\bjtitle{\jgr}
\bvolume{100},
\bfpage{3355}.
\doiurl{10.1029/94JA02591}.
\adsurl{1995JGR...100.3355F}.
\end{barticle}
\endbibitem

\bibitem[\protect\citeauthoryear{{Forbes} and
  {Priest}}{1995}]{1995ApJ...446..377F}
\begin{barticle}
\bauthor{\bsnm{{Forbes}}, \binits{T.G.}},
\bauthor{\bsnm{{Priest}}, \binits{E.R.}}:
\byear{1995},
\batitle{{Photospheric Magnetic Field Evolution and Eruptive Flares}}.
\bjtitle{\apj}
\bvolume{446},
\bfpage{377}.
\doiurl{10.1086/175797}.
\adsurl{1995ApJ...446..377F}.
\end{barticle}
\endbibitem

\bibitem[\protect\citeauthoryear{{Gibson} and
  {Fan}}{2006}]{2006ApJ...637L..65G}
\begin{barticle}
\bauthor{\bsnm{{Gibson}}, \binits{S.E.}},
\bauthor{\bsnm{{Fan}}, \binits{Y.}}:
\byear{2006},
\batitle{{The Partial Expulsion of a Magnetic Flux Rope}}.
\bjtitle{\apjl}
\bvolume{637},
\bfpage{L65}.
\doiurl{10.1086/500452}.
\adsurl{http://esoads.eso.org/abs/2006ApJ...637L..65G}.
\end{barticle}
\endbibitem

\bibitem[\protect\citeauthoryear{{Green}, {Kliem}, and
  {Wallace}}{2011}]{2011A&A...526A...2G}
\begin{barticle}
\bauthor{\bsnm{{Green}}, \binits{L.M.}},
\bauthor{\bsnm{{Kliem}}, \binits{B.}},
\bauthor{\bsnm{{Wallace}}, \binits{A.J.}}:
\byear{2011},
\batitle{{Photospheric flux cancellation and associated flux rope formation and
  eruption}}.
\bjtitle{\aap}
\bvolume{526},
\bfpage{A2}.
\doiurl{10.1051/0004-6361/201015146}.
\adsurl{2011A\%26A...526A...2G}.
\end{barticle}
\endbibitem

\bibitem[\protect\citeauthoryear{{Hagyard}, {Venkatakrishnan}, and
  {Smith}}{1990}]{1990ApJS...73..159H}
\begin{barticle}
\bauthor{\bsnm{{Hagyard}}, \binits{M.J.}},
\bauthor{\bsnm{{Venkatakrishnan}}, \binits{P.}},
\bauthor{\bsnm{{Smith}}, \binits{J.B.} \bsuffix{Jr.}}:
\byear{1990},
\batitle{{Nonpotential magnetic fields at sites of gamma-ray flares}}.
\bjtitle{\apjs}
\bvolume{73},
\bfpage{159}.
\doiurl{10.1086/191447}.
\adsurl{1990ApJS...73..159H}.
\end{barticle}
\endbibitem

\bibitem[\protect\citeauthoryear{{Harvey}
  \textit{et~al.}}{1996}]{1996Sci...272.1284H}
\begin{barticle}
\bauthor{\bsnm{{Harvey}}, \binits{J.W.}},
\bauthor{\bsnm{{Hill}}, \binits{F.}},
\bauthor{\bsnm{{Hubbard}}, \binits{R.P.}},
\bauthor{\bsnm{{Kennedy}}, \binits{J.R.}},
\bauthor{\bsnm{{Leibacher}}, \binits{J.W.}},
\bauthor{\bsnm{{Pintar}}, \binits{J.A.}},
\bauthor{\bsnm{{Gilman}}, \binits{P.A.}},
\bauthor{\bsnm{{Noyes}}, \binits{R.W.}},
\bauthor{\bsnm{{Title}}, \binits{A.M.}},
\bauthor{\bsnm{{Toomre}}, \binits{J.}},
\bauthor{\bsnm{{Ulrich}}, \binits{R.K.}},
\bauthor{\bsnm{{Bhatnagar}}, \binits{A.}},
\bauthor{\bsnm{{Kennewell}}, \binits{J.A.}},
\bauthor{\bsnm{{Marquette}}, \binits{W.}},
\bauthor{\bsnm{{Patron}}, \binits{J.}},
\bauthor{\bsnm{{Saa}}, \binits{O.}},
\bauthor{\bsnm{{Yasukawa}}, \binits{E.}}:
\byear{1996},
\batitle{{The Global Oscillation Network Group (GONG) Project}}.
\bjtitle{Science}
\bvolume{272},
\bfpage{1284}.
\doiurl{10.1126/science.272.5266.1284}.
\adsurl{1996Sci...272.1284H}.
\end{barticle}
\endbibitem

\bibitem[\protect\citeauthoryear{{Harvey}
  \textit{et~al.}}{2011}]{2011SPD....42.1745H}
\begin{bchapter}
\bauthor{\bsnm{{Harvey}}, \binits{J.W.}},
\bauthor{\bsnm{{Bolding}}, \binits{J.}},
\bauthor{\bsnm{{Clark}}, \binits{R.}},
\bauthor{\bsnm{{Hauth}}, \binits{D.}},
\bauthor{\bsnm{{Hill}}, \binits{F.}},
\bauthor{\bsnm{{Kroll}}, \binits{R.}},
\bauthor{\bsnm{{Luis}}, \binits{G.}},
\bauthor{\bsnm{{Mills}}, \binits{N.}},
\bauthor{\bsnm{{Purdy}}, \binits{T.}},
\bauthor{\bsnm{{Henney}}, \binits{C.}},
\bauthor{\bsnm{{Holland}}, \binits{D.}},
\bauthor{\bsnm{{Winter}}, \binits{J.}}:
\byear{2011},
\bctitle{{Full-disk Solar H-alpha Images From GONG}}.
In: \bbtitle{AAS/Solar Physics Division Abstracts \#42},
\bfpage{1745}.
\adsurl{2011SPD....42.1745H}.
\end{bchapter}
\endbibitem

\bibitem[\protect\citeauthoryear{{Howard}
  \textit{et~al.}}{2008}]{2008SSRv..136...67H}
\begin{barticle}
\bauthor{\bsnm{{Howard}}, \binits{R.A.}},
\bauthor{\bsnm{{Moses}}, \binits{J.D.}},
\bauthor{\bsnm{{Vourlidas}}, \binits{A.}},
\bauthor{\bsnm{{Newmark}}, \binits{J.S.}},
\bauthor{\bsnm{{Socker}}, \binits{D.G.}},
\bauthor{\bsnm{{Plunkett}}, \binits{S.P.}},
\bauthor{\bsnm{{Korendyke}}, \binits{C.M.}},
\bauthor{\bsnm{{Cook}}, \binits{J.W.}},
\bauthor{\bsnm{{Hurley}}, \binits{A.}},
\bauthor{\bsnm{{Davila}}, \binits{J.M.}},
\bauthor{\bsnm{{Thompson}}, \binits{W.T.}},
\bauthor{\bsnm{{St Cyr}}, \binits{O.C.}},
\bauthor{\bsnm{{Mentzell}}, \binits{E.}},
\bauthor{\bsnm{{Mehalick}}, \binits{K.}},
\bauthor{\bsnm{{Lemen}}, \binits{J.R.}},
\bauthor{\bsnm{{Wuelser}}, \binits{J.P.}},
\bauthor{\bsnm{{Duncan}}, \binits{D.W.}},
\bauthor{\bsnm{{Tarbell}}, \binits{T.D.}},
\bauthor{\bsnm{{Wolfson}}, \binits{C.J.}},
\bauthor{\bsnm{{Moore}}, \binits{A.}},
\bauthor{\bsnm{{Harrison}}, \binits{R.A.}},
\bauthor{\bsnm{{Waltham}}, \binits{N.R.}},
\bauthor{\bsnm{{Lang}}, \binits{J.}},
\bauthor{\bsnm{{Davis}}, \binits{C.J.}},
\bauthor{\bsnm{{Eyles}}, \binits{C.J.}},
\bauthor{\bsnm{{Mapson-Menard}}, \binits{H.}},
\bauthor{\bsnm{{Simnett}}, \binits{G.M.}},
\bauthor{\bsnm{{Halain}}, \binits{J.P.}},
\bauthor{\bsnm{{Defise}}, \binits{J.M.}},
\bauthor{\bsnm{{Mazy}}, \binits{E.}},
\bauthor{\bsnm{{Rochus}}, \binits{P.}},
\bauthor{\bsnm{{Mercier}}, \binits{R.}},
\bauthor{\bsnm{{Ravet}}, \binits{M.F.}},
\bauthor{\bsnm{{Delmotte}}, \binits{F.}},
\bauthor{\bsnm{{Auchere}}, \binits{F.}},
\bauthor{\bsnm{{Delaboudiniere}}, \binits{J.P.}},
\bauthor{\bsnm{{Bothmer}}, \binits{V.}},
\bauthor{\bsnm{{Deutsch}}, \binits{W.}},
\bauthor{\bsnm{{Wang}}, \binits{D.}},
\bauthor{\bsnm{{Rich}}, \binits{N.}},
\bauthor{\bsnm{{Cooper}}, \binits{S.}},
\bauthor{\bsnm{{Stephens}}, \binits{V.}},
\bauthor{\bsnm{{Maahs}}, \binits{G.}},
\bauthor{\bsnm{{Baugh}}, \binits{R.}},
\bauthor{\bsnm{{McMullin}}, \binits{D.}},
\bauthor{\bsnm{{Carter}}, \binits{T.}}:
\byear{2008},
\batitle{{Sun Earth Connection Coronal and Heliospheric Investigation
  (SECCHI)}}.
\bjtitle{\ssr}
\bvolume{136},
\bfpage{67}.
\doiurl{10.1007/s11214-008-9341-4}.
\adsurl{http://ads.ari.uni-heidelberg.de/abs/2008SSRv..136...67H}.
\end{barticle}
\endbibitem

\bibitem[\protect\citeauthoryear{{Kaneko} and
  {Yokoyama}}{2014}]{2014ApJ...796...44K}
\begin{barticle}
\bauthor{\bsnm{{Kaneko}}, \binits{T.}},
\bauthor{\bsnm{{Yokoyama}}, \binits{T.}}:
\byear{2014},
\batitle{{Simulation Study of Solar Plasma Eruptions Caused by Interactions
  between Emerging Flux and Coronal Arcade Fields}}.
\bjtitle{\apj}
\bvolume{796},
\bfpage{44}.
\doiurl{10.1088/0004-637X/796/1/44}.
\adsurl{2014ApJ...796...44K}.
\end{barticle}
\endbibitem

\bibitem[\protect\citeauthoryear{{Karpen}, {Antiochos}, and
  {DeVore}}{2012}]{2012ApJ...760...81K}
\begin{barticle}
\bauthor{\bsnm{{Karpen}}, \binits{J.T.}},
\bauthor{\bsnm{{Antiochos}}, \binits{S.K.}},
\bauthor{\bsnm{{DeVore}}, \binits{C.R.}}:
\byear{2012},
\batitle{{The Mechanisms for the Onset and Explosive Eruption of Coronal Mass
  Ejections and Eruptive Flares}}.
\bjtitle{\apj}
\bvolume{760},
\bfpage{81}.
\doiurl{10.1088/0004-637X/760/1/81}.
\adsurl{2012ApJ...760...81K}.
\end{barticle}
\endbibitem

\bibitem[\protect\citeauthoryear{{Kliem} and
  {T{\"o}r{\"o}k}}{2006}]{2006PhRvL..96y5002K}
\begin{barticle}
\bauthor{\bsnm{{Kliem}}, \binits{B.}},
\bauthor{\bsnm{{T{\"o}r{\"o}k}}, \binits{T.}}:
\byear{2006},
\batitle{{Torus Instability}}.
\bjtitle{\PhRvL}
\bvolume{96}(\bissue{25}),
\bfpage{255002}.
\doiurl{10.1103/PhysRevLett.96.255002}.
\adsurl{2006PhRvL..96y5002K}.
\end{barticle}
\endbibitem

\bibitem[\protect\citeauthoryear{{Kusano}
  \textit{et~al.}}{2012}]{2012ApJ...760...31K}
\begin{barticle}
\bauthor{\bsnm{{Kusano}}, \binits{K.}},
\bauthor{\bsnm{{Bamba}}, \binits{Y.}},
\bauthor{\bsnm{{Yamamoto}}, \binits{T.T.}},
\bauthor{\bsnm{{Iida}}, \binits{Y.}},
\bauthor{\bsnm{{Toriumi}}, \binits{S.}},
\bauthor{\bsnm{{Asai}}, \binits{A.}}:
\byear{2012},
\batitle{{Magnetic Field Structures Triggering Solar Flares and Coronal Mass
  Ejections}}.
\bjtitle{\apj}
\bvolume{760},
\bfpage{31}.
\doiurl{10.1088/0004-637X/760/1/31}.
\adsurl{http://ads.ari.uni-heidelberg.de/abs/2012ApJ...760...31K}.
\end{barticle}
\endbibitem

\bibitem[\protect\citeauthoryear{{Leake}, {Linton}, and
  {Antiochos}}{2014}]{2014ApJ...787...46L}
\begin{barticle}
\bauthor{\bsnm{{Leake}}, \binits{J.E.}},
\bauthor{\bsnm{{Linton}}, \binits{M.G.}},
\bauthor{\bsnm{{Antiochos}}, \binits{S.K.}}:
\byear{2014},
\batitle{{Simulations of Emerging Magnetic Flux. II. The Formation of Unstable
  Coronal Flux Ropes and the Initiation of Coronal Mass Ejections}}.
\bjtitle{\apj}
\bvolume{787},
\bfpage{46}.
\doiurl{10.1088/0004-637X/787/1/46}.
\adsurl{2014ApJ...787...46L}.
\end{barticle}
\endbibitem

\bibitem[\protect\citeauthoryear{{Lemen}
  \textit{et~al.}}{2012}]{2012SoPh..275...17L}
\begin{barticle}
\bauthor{\bsnm{{Lemen}}, \binits{J.R.}},
\bauthor{\bsnm{{Title}}, \binits{A.M.}},
\bauthor{\bsnm{{Akin}}, \binits{D.J.}},
\bauthor{\bsnm{{Boerner}}, \binits{P.F.}},
\bauthor{\bsnm{{Chou}}, \binits{C.}},
\bauthor{\bsnm{{Drake}}, \binits{J.F.}},
\bauthor{\bsnm{{Duncan}}, \binits{D.W.}},
\bauthor{\bsnm{{Edwards}}, \binits{C.G.}},
\bauthor{\bsnm{{Friedlaender}}, \binits{F.M.}},
\bauthor{\bsnm{{Heyman}}, \binits{G.F.}},
\bauthor{\bsnm{{Hurlburt}}, \binits{N.E.}},
\bauthor{\bsnm{{Katz}}, \binits{N.L.}},
\bauthor{\bsnm{{Kushner}}, \binits{G.D.}},
\bauthor{\bsnm{{Levay}}, \binits{M.}},
\bauthor{\bsnm{{Lindgren}}, \binits{R.W.}},
\bauthor{\bsnm{{Mathur}}, \binits{D.P.}},
\bauthor{\bsnm{{McFeaters}}, \binits{E.L.}},
\bauthor{\bsnm{{Mitchell}}, \binits{S.}},
\bauthor{\bsnm{{Rehse}}, \binits{R.A.}},
\bauthor{\bsnm{{Schrijver}}, \binits{C.J.}},
\bauthor{\bsnm{{Springer}}, \binits{L.A.}},
\bauthor{\bsnm{{Stern}}, \binits{R.A.}},
\bauthor{\bsnm{{Tarbell}}, \binits{T.D.}},
\bauthor{\bsnm{{Wuelser}}, \binits{J.-P.}},
\bauthor{\bsnm{{Wolfson}}, \binits{C.J.}},
\bauthor{\bsnm{{Yanari}}, \binits{C.}},
\bauthor{\bsnm{{Bookbinder}}, \binits{J.A.}},
\bauthor{\bsnm{{Cheimets}}, \binits{P.N.}},
\bauthor{\bsnm{{Caldwell}}, \binits{D.}},
\bauthor{\bsnm{{Deluca}}, \binits{E.E.}},
\bauthor{\bsnm{{Gates}}, \binits{R.}},
\bauthor{\bsnm{{Golub}}, \binits{L.}},
\bauthor{\bsnm{{Park}}, \binits{S.}},
\bauthor{\bsnm{{Podgorski}}, \binits{W.A.}},
\bauthor{\bsnm{{Bush}}, \binits{R.I.}},
\bauthor{\bsnm{{Scherrer}}, \binits{P.H.}},
\bauthor{\bsnm{{Gummin}}, \binits{M.A.}},
\bauthor{\bsnm{{Smith}}, \binits{P.}},
\bauthor{\bsnm{{Auker}}, \binits{G.}},
\bauthor{\bsnm{{Jerram}}, \binits{P.}},
\bauthor{\bsnm{{Pool}}, \binits{P.}},
\bauthor{\bsnm{{Soufli}}, \binits{R.}},
\bauthor{\bsnm{{Windt}}, \binits{D.L.}},
\bauthor{\bsnm{{Beardsley}}, \binits{S.}},
\bauthor{\bsnm{{Clapp}}, \binits{M.}},
\bauthor{\bsnm{{Lang}}, \binits{J.}},
\bauthor{\bsnm{{Waltham}}, \binits{N.}}:
\byear{2012},
\batitle{{The Atmospheric Imaging Assembly (AIA) on the Solar Dynamics
  Observatory (SDO)}}.
\bjtitle{\solphys}
\bvolume{275},
\bfpage{17}.
\doiurl{10.1007/s11207-011-9776-8}.
\adsurl{2012SoPh..275...17L}.
\end{barticle}
\endbibitem

\bibitem[\protect\citeauthoryear{{Li}
  \textit{et~al.}}{2000}]{2000PASJ...52..465L}
\begin{barticle}
\bauthor{\bsnm{{Li}}, \binits{H.}},
\bauthor{\bsnm{{Sakurai}}, \binits{T.}},
\bauthor{\bsnm{{Ichimoto}}, \binits{K.}},
\bauthor{\bsnm{{UeNo}}, \binits{S.}}:
\byear{2000},
\batitle{{Magnetic Field Evolution Leading to Solar Flares I. Cases with Low
  Magnetic Shear and Flux Emergence}}.
\bjtitle{\pasj}
\bvolume{52},
\bfpage{465}.
\adsurl{2000PASJ...52..465L}.
\end{barticle}
\endbibitem

\bibitem[\protect\citeauthoryear{{Lin} and
  {Forbes}}{2000}]{2000JGR...105.2375L}
\begin{barticle}
\bauthor{\bsnm{{Lin}}, \binits{J.}},
\bauthor{\bsnm{{Forbes}}, \binits{T.G.}}:
\byear{2000},
\batitle{{Effects of reconnection on the coronal mass ejection process}}.
\bjtitle{\jgr}
\bvolume{105},
\bfpage{2375}.
\doiurl{10.1029/1999JA900477}.
\adsurl{2000JGR...105.2375L}.
\end{barticle}
\endbibitem

\bibitem[\protect\citeauthoryear{{Lin}, {Forbes}, and
  {Isenberg}}{2001}]{2001JGR...10625053L}
\begin{barticle}
\bauthor{\bsnm{{Lin}}, \binits{J.}},
\bauthor{\bsnm{{Forbes}}, \binits{T.G.}},
\bauthor{\bsnm{{Isenberg}}, \binits{P.A.}}:
\byear{2001},
\batitle{{Prominence eruptions and coronal mass ejections triggered by newly
  emerging flux}}.
\bjtitle{\jgr}
\bvolume{106},
\bfpage{25053}.
\doiurl{10.1029/2001JA000046}.
\adsurl{2001JGR...10625053L}.
\end{barticle}
\endbibitem

\bibitem[\protect\citeauthoryear{{Lites}}{2005}]{2005ApJ...622.1275L}
\begin{barticle}
\bauthor{\bsnm{{Lites}}, \binits{B.W.}}:
\byear{2005},
\batitle{{Magnetic Flux Ropes in the Solar Photosphere: The Vector Magnetic
  Field under Active Region Filaments}}.
\bjtitle{\apj}
\bvolume{622},
\bfpage{1275}.
\doiurl{10.1086/428080}.
\adsurl{2005ApJ...622.1275L}.
\end{barticle}
\endbibitem

\bibitem[\protect\citeauthoryear{{Livi}
  \textit{et~al.}}{1989}]{1989SoPh..121..197L}
\begin{barticle}
\bauthor{\bsnm{{Livi}}, \binits{S.H.B.}},
\bauthor{\bsnm{{Martin}}, \binits{S.}},
\bauthor{\bsnm{{Wang}}, \binits{H.}},
\bauthor{\bsnm{{Ai}}, \binits{G.}}:
\byear{1989},
\batitle{{The association of flares to cancelling magnetic features on the
  sun}}.
\bjtitle{\solphys}
\bvolume{121},
\bfpage{197}.
\doiurl{10.1007/BF00161696}.
\adsurl{1989SoPh..121..197L}.
\end{barticle}
\endbibitem

\bibitem[\protect\citeauthoryear{{Louis}
  \textit{et~al.}}{2014}]{2014A&A...562A.110L}
\begin{barticle}
\bauthor{\bsnm{{Louis}}, \binits{R.E.}},
\bauthor{\bsnm{{Puschmann}}, \binits{K.G.}},
\bauthor{\bsnm{{Kliem}}, \binits{B.}},
\bauthor{\bsnm{{Balthasar}}, \binits{H.}},
\bauthor{\bsnm{{Denker}}, \binits{C.}}:
\byear{2014},
\batitle{{Sunspot splitting triggering an eruptive flare}}.
\bjtitle{\aap}
\bvolume{562},
\bfpage{A110}.
\doiurl{10.1051/0004-6361/201321106}.
\adsurl{2014A\%26A...562A.110L}.
\end{barticle}
\endbibitem

\bibitem[\protect\citeauthoryear{{Low}}{1996}]{1996SoPh..167..217L}
\begin{barticle}
\bauthor{\bsnm{{Low}}, \binits{B.C.}}:
\byear{1996},
\batitle{{Solar Activity and the Corona}}.
\bjtitle{\solphys}
\bvolume{167},
\bfpage{217}.
\doiurl{10.1007/BF00146338}.
\adsurl{1996SoPh..167..217L}.
\end{barticle}
\endbibitem

\bibitem[\protect\citeauthoryear{{Lynch}
  \textit{et~al.}}{2008}]{2008ApJ...683.1192L}
\begin{barticle}
\bauthor{\bsnm{{Lynch}}, \binits{B.J.}},
\bauthor{\bsnm{{Antiochos}}, \binits{S.K.}},
\bauthor{\bsnm{{DeVore}}, \binits{C.R.}},
\bauthor{\bsnm{{Luhmann}}, \binits{J.G.}},
\bauthor{\bsnm{{Zurbuchen}}, \binits{T.H.}}:
\byear{2008},
\batitle{{Topological Evolution of a Fast Magnetic Breakout CME in Three
  Dimensions}}.
\bjtitle{\apj}
\bvolume{683},
\bfpage{1192}.
\doiurl{10.1086/589738}.
\adsurl{http://esoads.eso.org/abs/2008ApJ...683.1192L}.
\end{barticle}
\endbibitem

\bibitem[\protect\citeauthoryear{{Mackay} and {van
  Ballegooijen}}{2006}]{2006ApJ...641..577M}
\begin{barticle}
\bauthor{\bsnm{{Mackay}}, \binits{D.H.}},
\bauthor{\bsnm{{van Ballegooijen}}, \binits{A.A.}}:
\byear{2006},
\batitle{{Models of the Large-Scale Corona. I. Formation, Evolution, and
  Liftoff of Magnetic Flux Ropes}}.
\bjtitle{\apj}
\bvolume{641},
\bfpage{577}.
\doiurl{10.1086/500425}.
\adsurl{2006ApJ...641..577M}.
\end{barticle}
\endbibitem

\bibitem[\protect\citeauthoryear{{Mackay}
  \textit{et~al.}}{2010}]{2010SSRv..151..333M}
\begin{barticle}
\bauthor{\bsnm{{Mackay}}, \binits{D.H.}},
\bauthor{\bsnm{{Karpen}}, \binits{J.T.}},
\bauthor{\bsnm{{Ballester}}, \binits{J.L.}},
\bauthor{\bsnm{{Schmieder}}, \binits{B.}},
\bauthor{\bsnm{{Aulanier}}, \binits{G.}}:
\byear{2010},
\batitle{{Physics of Solar Prominences: II---Magnetic Structure and Dynamics}}.
\bjtitle{\ssr}
\bvolume{151},
\bfpage{333}.
\doiurl{10.1007/s11214-010-9628-0}.
\adsurl{2010SSRv..151..333M}.
\end{barticle}
\endbibitem

\bibitem[\protect\citeauthoryear{{Manchester}
  \textit{et~al.}}{2004}]{2004ApJ...610..588M}
\begin{barticle}
\bauthor{\bsnm{{Manchester}}, \binits{W.} \bsuffix{IV}},
\bauthor{\bsnm{{Gombosi}}, \binits{T.}},
\bauthor{\bsnm{{DeZeeuw}}, \binits{D.}},
\bauthor{\bsnm{{Fan}}, \binits{Y.}}:
\byear{2004},
\batitle{{Eruption of a Buoyantly Emerging Magnetic Flux Rope}}.
\bjtitle{\apj}
\bvolume{610},
\bfpage{588}.
\doiurl{10.1086/421516}.
\adsurl{2004ApJ...610..588M}.
\end{barticle}
\endbibitem

\bibitem[\protect\citeauthoryear{{Martin}}{1998}]{1998SoPh..182..107M}
\begin{barticle}
\bauthor{\bsnm{{Martin}}, \binits{S.F.}}:
\byear{1998},
\batitle{{Conditions for the Formation and Maintenance of Filaments (Invited
  Review)}}.
\bjtitle{\solphys}
\bvolume{182},
\bfpage{107}.
\doiurl{10.1023/A:1005026814076}.
\adsurl{1998SoPh..182..107M}.
\end{barticle}
\endbibitem

\bibitem[\protect\citeauthoryear{{Martin}
  \textit{et~al.}}{1982}]{1982AdSpR...2...39M}
\begin{barticle}
\bauthor{\bsnm{{Martin}}, \binits{S.F.}},
\bauthor{\bsnm{{Dezso}}, \binits{L.}},
\bauthor{\bsnm{{Antalova}}, \binits{A.}},
\bauthor{\bsnm{{Kucera}}, \binits{A.}},
\bauthor{\bsnm{{Harvey}}, \binits{K.L.}}:
\byear{1982},
\batitle{{Emerging magnetic flux, flares and filaments - FBS interval 16-23
  June 1980}}.
\bjtitle{Advances in Space Research}
\bvolume{2},
\bfpage{39}.
\doiurl{10.1016/0273-1177(82)90177-6}.
\adsurl{1982AdSpR...2...39M}.
\end{barticle}
\endbibitem

\bibitem[\protect\citeauthoryear{{Martres}
  \textit{et~al.}}{1968}]{1968SoPh....5..187M}
\begin{barticle}
\bauthor{\bsnm{{Martres}}, \binits{M.-J.}},
\bauthor{\bsnm{{Michard}}, \binits{R.}},
\bauthor{\bsnm{{Soru-Iscovici}}, \binits{I.}},
\bauthor{\bsnm{{Tsap}}, \binits{T.T.}}:
\byear{1968},
\batitle{{\'Etude de la localisation des \'eruptions dans la structure
  magn\'etique \'evolutive des r\'egions actives solaires}}.
\bjtitle{\solphys}
\bvolume{5},
\bfpage{187}.
\doiurl{10.1007/BF00147965}.
\adsurl{1968SoPh....5..187M}.
\end{barticle}
\endbibitem

\bibitem[\protect\citeauthoryear{{Mikic} and
  {Linker}}{1994}]{1994ApJ...430..898M}
\begin{barticle}
\bauthor{\bsnm{{Mikic}}, \binits{Z.}},
\bauthor{\bsnm{{Linker}}, \binits{J.A.}}:
\byear{1994},
\batitle{{Disruption of coronal magnetic field arcades}}.
\bjtitle{\apj}
\bvolume{430},
\bfpage{898}.
\doiurl{10.1086/174460}.
\adsurl{http://ads.ari.uni-heidelberg.de/abs/1994ApJ...430..898M}.
\end{barticle}
\endbibitem

\bibitem[\protect\citeauthoryear{{Pesnell}, {Thompson}, and
  {Chamberlin}}{2012}]{2012SoPh..275....3P}
\begin{barticle}
\bauthor{\bsnm{{Pesnell}}, \binits{W.D.}},
\bauthor{\bsnm{{Thompson}}, \binits{B.J.}},
\bauthor{\bsnm{{Chamberlin}}, \binits{P.C.}}:
\byear{2012},
\batitle{{The Solar Dynamics Observatory (SDO)}}.
\bjtitle{\solphys}
\bvolume{275},
\bfpage{3}.
\doiurl{10.1007/s11207-011-9841-3}.
\adsurl{2012SoPh..275....3P}.
\end{barticle}
\endbibitem

\bibitem[\protect\citeauthoryear{{Roumeliotis}, {Sturrock}, and
  {Antiochos}}{1994}]{1994ApJ...423..847R}
\begin{barticle}
\bauthor{\bsnm{{Roumeliotis}}, \binits{G.}},
\bauthor{\bsnm{{Sturrock}}, \binits{P.A.}},
\bauthor{\bsnm{{Antiochos}}, \binits{S.K.}}:
\byear{1994},
\batitle{{A Numerical Study of the Sudden Eruption of Sheared Magnetic
  Fields}}.
\bjtitle{\apj}
\bvolume{423},
\bfpage{847}.
\doiurl{10.1086/173862}.
\adsurl{http://ads.ari.uni-heidelberg.de/abs/1994ApJ...423..847R}.
\end{barticle}
\endbibitem

\bibitem[\protect\citeauthoryear{{Rust}}{1972}]{1972SoPh...25..141R}
\begin{barticle}
\bauthor{\bsnm{{Rust}}, \binits{D.M.}}:
\byear{1972},
\batitle{{Flares and Changing Magnetic Fields}}.
\bjtitle{\solphys}
\bvolume{25},
\bfpage{141}.
\doiurl{10.1007/BF00155753}.
\adsurl{1972SoPh...25..141R}.
\end{barticle}
\endbibitem

\bibitem[\protect\citeauthoryear{{Sakajiri}
  \textit{et~al.}}{2004}]{2004ApJ...616..578S}
\begin{barticle}
\bauthor{\bsnm{{Sakajiri}}, \binits{T.}},
\bauthor{\bsnm{{Brooks}}, \binits{D.H.}},
\bauthor{\bsnm{{Yamamoto}}, \binits{T.}},
\bauthor{\bsnm{{Shiota}}, \binits{D.}},
\bauthor{\bsnm{{Isobe}}, \binits{H.}},
\bauthor{\bsnm{{Akiyama}}, \binits{S.}},
\bauthor{\bsnm{{Ueno}}, \binits{S.}},
\bauthor{\bsnm{{Kitai}}, \binits{R.}},
\bauthor{\bsnm{{Shibata}}, \binits{K.}}:
\byear{2004},
\batitle{{A Study of a Tiny Two-Ribbon Flare Driven by Emerging Flux}}.
\bjtitle{\apj}
\bvolume{616},
\bfpage{578}.
\doiurl{10.1086/424823}.
\adsurl{2004ApJ...616..578S}.
\end{barticle}
\endbibitem

\bibitem[\protect\citeauthoryear{{Savcheva}
  \textit{et~al.}}{2012}]{2012ApJ...759..105S}
\begin{barticle}
\bauthor{\bsnm{{Savcheva}}, \binits{A.S.}},
\bauthor{\bsnm{{Green}}, \binits{L.M.}},
\bauthor{\bsnm{{van Ballegooijen}}, \binits{A.A.}},
\bauthor{\bsnm{{DeLuca}}, \binits{E.E.}}:
\byear{2012},
\batitle{{Photospheric Flux Cancellation and the Build-up of Sigmoidal Flux
  Ropes on the Sun}}.
\bjtitle{\apj}
\bvolume{759},
\bfpage{105}.
\doiurl{10.1088/0004-637X/759/2/105}.
\adsurl{2012ApJ...759..105S}.
\end{barticle}
\endbibitem

\bibitem[\protect\citeauthoryear{{Schou}
  \textit{et~al.}}{2012}]{2012SoPh..275..229S}
\begin{barticle}
\bauthor{\bsnm{{Schou}}, \binits{J.}},
\bauthor{\bsnm{{Scherrer}}, \binits{P.H.}},
\bauthor{\bsnm{{Bush}}, \binits{R.I.}},
\bauthor{\bsnm{{Wachter}}, \binits{R.}},
\bauthor{\bsnm{{Couvidat}}, \binits{S.}},
\bauthor{\bsnm{{Rabello-Soares}}, \binits{M.C.}},
\bauthor{\bsnm{{Bogart}}, \binits{R.S.}},
\bauthor{\bsnm{{Hoeksema}}, \binits{J.T.}},
\bauthor{\bsnm{{Liu}}, \binits{Y.}},
\bauthor{\bsnm{{Duvall}}, \binits{T.L.}},
\bauthor{\bsnm{{Akin}}, \binits{D.J.}},
\bauthor{\bsnm{{Allard}}, \binits{B.A.}},
\bauthor{\bsnm{{Miles}}, \binits{J.W.}},
\bauthor{\bsnm{{Rairden}}, \binits{R.}},
\bauthor{\bsnm{{Shine}}, \binits{R.A.}},
\bauthor{\bsnm{{Tarbell}}, \binits{T.D.}},
\bauthor{\bsnm{{Title}}, \binits{A.M.}},
\bauthor{\bsnm{{Wolfson}}, \binits{C.J.}},
\bauthor{\bsnm{{Elmore}}, \binits{D.F.}},
\bauthor{\bsnm{{Norton}}, \binits{A.A.}},
\bauthor{\bsnm{{Tomczyk}}, \binits{S.}}:
\byear{2012},
\batitle{{Design and Ground Calibration of the Helioseismic and Magnetic Imager
  (HMI) Instrument on the Solar Dynamics Observatory (SDO)}}.
\bjtitle{\solphys}
\bvolume{275},
\bfpage{229}.
\doiurl{10.1007/s11207-011-9842-2}.
\adsurl{2012SoPh..275..229S}.
\end{barticle}
\endbibitem

\bibitem[\protect\citeauthoryear{{Schrijver}}{2009}]{2009AdSpR..43..739S}
\begin{barticle}
\bauthor{\bsnm{{Schrijver}}, \binits{C.J.}}:
\byear{2009},
\batitle{{Driving major solar flares and eruptions: A review}}.
\bjtitle{Advances in Space Research}
\bvolume{43},
\bfpage{739}.
\doiurl{10.1016/j.asr.2008.11.004}.
\adsurl{2009AdSpR..43..739S}.
\end{barticle}
\endbibitem

\bibitem[\protect\citeauthoryear{{Schuck}}{2006}]{2006ApJ...646.1358S}
\begin{barticle}
\bauthor{\bsnm{{Schuck}}, \binits{P.W.}}:
\byear{2006},
\batitle{{Tracking Magnetic Footpoints with the Magnetic Induction Equation}}.
\bjtitle{\apj}
\bvolume{646},
\bfpage{1358}.
\doiurl{10.1086/505015}.
\adsurl{2006ApJ...646.1358S}.
\end{barticle}
\endbibitem

\bibitem[\protect\citeauthoryear{{Sterling}
  \textit{et~al.}}{2010}]{2010A&A...521A..49S}
\begin{barticle}
\bauthor{\bsnm{{Sterling}}, \binits{A.C.}},
\bauthor{\bsnm{{Chifor}}, \binits{C.}},
\bauthor{\bsnm{{Mason}}, \binits{H.E.}},
\bauthor{\bsnm{{Moore}}, \binits{R.L.}},
\bauthor{\bsnm{{Young}}, \binits{P.R.}}:
\byear{2010},
\batitle{{Evidence for magnetic flux cancelation leading to an ejective solar
  eruption observed by Hinode, TRACE, STEREO, and SoHO/MDI}}.
\bjtitle{\aap}
\bvolume{521},
\bfpage{A49}.
\doiurl{10.1051/0004-6361/201014006}.
\adsurl{2010A\%26A...521A..49S}.
\end{barticle}
\endbibitem

\bibitem[\protect\citeauthoryear{{Sun}
  \textit{et~al.}}{2012}]{2012ApJ...748...77S}
\begin{barticle}
\bauthor{\bsnm{{Sun}}, \binits{X.}},
\bauthor{\bsnm{{Hoeksema}}, \binits{J.T.}},
\bauthor{\bsnm{{Liu}}, \binits{Y.}},
\bauthor{\bsnm{{Wiegelmann}}, \binits{T.}},
\bauthor{\bsnm{{Hayashi}}, \binits{K.}},
\bauthor{\bsnm{{Chen}}, \binits{Q.}},
\bauthor{\bsnm{{Thalmann}}, \binits{J.}}:
\byear{2012},
\batitle{{Evolution of Magnetic Field and Energy in a Major Eruptive Active
  Region Based on SDO/HMI Observation}}.
\bjtitle{\apj}
\bvolume{748},
\bfpage{77}.
\doiurl{10.1088/0004-637X/748/2/77}.
\adsurl{2012ApJ...748...77S}.
\end{barticle}
\endbibitem

\bibitem[\protect\citeauthoryear{{Titov} and
  {D{\'e}moulin}}{1999}]{Titov&Demoulin1999}
\begin{barticle}
\bauthor{\bsnm{{Titov}}, \binits{V.S.}},
\bauthor{\bsnm{{D{\'e}moulin}}, \binits{P.}}:
\byear{1999},
\batitle{{Basic topology of twisted magnetic configurations in solar flares}}.
\bjtitle{\aap}
\bvolume{351},
\bfpage{707}.
\adsurl{1999A\%26A...351..707T}.
\end{barticle}
\endbibitem

\bibitem[\protect\citeauthoryear{{T{\"o}r{\"o}k} and
  {Kliem}}{2003}]{Torok&Kliem2003}
\begin{barticle}
\bauthor{\bsnm{{T{\"o}r{\"o}k}}, \binits{T.}},
\bauthor{\bsnm{{Kliem}}, \binits{B.}}:
\byear{2003},
\batitle{{The evolution of twisting coronal magnetic flux tubes}}.
\bjtitle{\aap}
\bvolume{406},
\bfpage{1043}.
\doiurl{10.1051/0004-6361:20030692}.
\adsurl{http://ads.ari.uni-heidelberg.de/abs/2003A\%26A...406.1043T}.
\end{barticle}
\endbibitem

\bibitem[\protect\citeauthoryear{{van Ballegooijen} and
  {Martens}}{1989}]{1989ApJ...343..971V}
\begin{barticle}
\bauthor{\bsnm{{van Ballegooijen}}, \binits{A.A.}},
\bauthor{\bsnm{{Martens}}, \binits{P.C.H.}}:
\byear{1989},
\batitle{{Formation and eruption of solar prominences}}.
\bjtitle{\apj}
\bvolume{343},
\bfpage{971}.
\doiurl{10.1086/167766}.
\adsurl{1989ApJ...343..971V}.
\end{barticle}
\endbibitem

\bibitem[\protect\citeauthoryear{{van Tend} and
  {Kuperus}}{1978}]{1978SoPh...59..115V}
\begin{barticle}
\bauthor{\bsnm{{van Tend}}, \binits{W.}},
\bauthor{\bsnm{{Kuperus}}, \binits{M.}}:
\byear{1978},
\batitle{{The development of coronal electric current systems in active regions
  and their relation to filaments and flares}}.
\bjtitle{\solphys}
\bvolume{59},
\bfpage{115}.
\doiurl{10.1007/BF00154935}.
\adsurl{http://esoads.eso.org/abs/1978SoPh...59..115V}.
\end{barticle}
\endbibitem

\bibitem[\protect\citeauthoryear{{Wang}}{2006}]{2006ApJ...649..490W}
\begin{barticle}
\bauthor{\bsnm{{Wang}}, \binits{H.}}:
\byear{2006},
\batitle{{Rapid Changes of Photospheric Magnetic Fields around Flaring Magnetic
  Neutral Lines}}.
\bjtitle{\apj}
\bvolume{649},
\bfpage{490}.
\doiurl{10.1086/506320}.
\adsurl{2006ApJ...649..490W}.
\end{barticle}
\endbibitem

\bibitem[\protect\citeauthoryear{{Wang} and {Shi}}{1993}]{1993SoPh..143..119W}
\begin{barticle}
\bauthor{\bsnm{{Wang}}, \binits{J.}},
\bauthor{\bsnm{{Shi}}, \binits{Z.}}:
\byear{1993},
\batitle{{The flare-associated magnetic changes in an active region. II - Flux
  emergence and cancellation}}.
\bjtitle{\solphys}
\bvolume{143},
\bfpage{119}.
\doiurl{10.1007/BF00619100}.
\adsurl{1993SoPh..143..119W}.
\end{barticle}
\endbibitem

\bibitem[\protect\citeauthoryear{{Wang} and
  {Sheeley}}{1999}]{1999ApJ...510L.157W}
\begin{barticle}
\bauthor{\bsnm{{Wang}}, \binits{Y.-M.}},
\bauthor{\bsnm{{Sheeley}}, \binits{N.R.} \bsuffix{Jr.}}:
\byear{1999},
\batitle{{Filament Eruptions near Emerging Bipoles}}.
\bjtitle{\apjl}
\bvolume{510},
\bfpage{L157}.
\doiurl{10.1086/311815}.
\adsurl{http://ads.ari.uni-heidelberg.de/abs/1999ApJ...510L.157W}.
\end{barticle}
\endbibitem

\bibitem[\protect\citeauthoryear{{Xu}, {Chen}, and
  {Fang}}{2005}]{2005ChJAA...5..636X}
\begin{barticle}
\bauthor{\bsnm{{Xu}}, \binits{X.-Y.}},
\bauthor{\bsnm{{Chen}}, \binits{P.-F.}},
\bauthor{\bsnm{{Fang}}, \binits{C.}}:
\byear{2005},
\batitle{{A Parametric Survey of the CME Triggering Process by Numerical
  Simulations}}.
\bjtitle{\cjaa}
\bvolume{5},
\bfpage{636}.
\doiurl{10.1088/1009-9271/5/6/010}.
\adsurl{http://ads.ari.uni-heidelberg.de/abs/2005ChJAA...5..636X}.
\end{barticle}
\endbibitem

\bibitem[\protect\citeauthoryear{{Xu}, {Fang}, and
  {Chen}}{2008}]{2008ChA&A..32...56X}
\begin{barticle}
\bauthor{\bsnm{{Xu}}, \binits{X.-Y.}},
\bauthor{\bsnm{{Fang}}, \binits{C.}},
\bauthor{\bsnm{{Chen}}, \binits{P.-F.}}:
\byear{2008},
\batitle{{A Statistical Study on the Filament Eruption Caused by New Emerging
  Flux}}.
\bjtitle{\caa}
\bvolume{32},
\bfpage{56}.
\doiurl{10.1016/j.chinastron.2008.01.008}.
\adsurl{http://ads.ari.uni-heidelberg.de/abs/2008ChA\%26A..32...56X}.
\end{barticle}
\endbibitem

\bibitem[\protect\citeauthoryear{{Zhang}, {Cheng}, and
  {Ding}}{2012}]{2012NatCo...3E.747Z}
\begin{barticle}
\bauthor{\bsnm{{Zhang}}, \binits{J.}},
\bauthor{\bsnm{{Cheng}}, \binits{X.}},
\bauthor{\bsnm{{Ding}}, \binits{M.-D.}}:
\byear{2012},
\batitle{{Observation of an evolving magnetic flux rope before and during a
  solar eruption}}.
\bjtitle{Nature Communications}
\bvolume{3},
\bfpage{747}.
\doiurl{10.1038/ncomms1753}.
\adsurl{http://esoads.eso.org/abs/2012NatCo...3E.747Z}.
\end{barticle}
\endbibitem

\bibitem[\protect\citeauthoryear{{Zhao} and
  {Hoeksema}}{1995}]{1995JGR...100...19Z}
\begin{barticle}
\bauthor{\bsnm{{Zhao}}, \binits{X.}},
\bauthor{\bsnm{{Hoeksema}}, \binits{J.T.}}:
\byear{1995},
\batitle{{Prediction of the interplanetary magnetic field strength}}.
\bjtitle{\jgr}
\bvolume{100},
\bfpage{19}.
\doiurl{10.1029/94JA02266}.
\adsurl{http://esoads.eso.org/abs/1995JGR...100...19Z}.
\end{barticle}
\endbibitem

\end{thebibliography}

\end{article}

\end{document}